\documentclass[prx,twocolumn,superscriptaddress,secnumarabic,balancelastpage,longbibliography]{revtex4-1}
\usepackage{siunitx}
\usepackage{braket}
\usepackage{amssymb}
\usepackage{amsmath,amsthm}
\usepackage[dvipsnames]{xcolor}
\usepackage{chngpage}
\usepackage{lgrind}        
\usepackage{color}         
\usepackage{graphics}      
\usepackage[pdftex]{graphicx}      
\usepackage{longtable}     
\usepackage{epsf}          
\usepackage{bm}            
\usepackage{asymptote}     
\usepackage{thumbpdf}
\usepackage{verbatim}
\usepackage{url}
\usepackage{MnSymbol}
\usepackage{physics}
\usepackage[colorlinks=true]{hyperref} 
\usepackage{enumitem}	

\usepackage{float}
\usepackage{times}
\usepackage{color}
\usepackage{verbatim}
\usepackage{soul,xcolor}
\setstcolor{red}
\usepackage{colortbl}
\usepackage[normalem]{ulem}

\usepackage{tabularx}
\usepackage{bbold}

\makeatletter
\AtBeginDocument{
\g@addto@macro{\appendix}{\renewcommand{\p@subsection}{\@Alph\c@section}}
}
\makeatother

\begin{document}
\newcommand\numberthis{\addtocounter{equation}{1,2}\tag{\theequation}}
\title{
Signal amplification in a solid-state quantum sensor 
via asymmetric time-reversal of many-body dynamics}

\affiliation{Department~of~Physics,~Harvard~University,~Cambridge,~MA~02138,~USA}
\affiliation{Materials~Department,~University~of~California,~Santa~Barbara,~CA~93106,~USA}
\affiliation{QuEra Computing Inc., Boston, MA 02135, USA}
\affiliation{Harvard~Quantum~Initiative,~Harvard~University,~Cambridge,~MA~02138,~USA}
\affiliation{Department~of~Physics,~University~of~California,~Santa~Barbara,~CA~93106,~USA}
\affiliation{Department~of~Chemistry~and~Chemical~Biology,~Harvard~University,~Cambridge,~MA~02138,~USA}

\author{Haoyang~Gao$^1$}
\thanks{These authors contributed equally to this work}
\author{Leigh~S.~Martin$^1$}
\thanks{These authors contributed equally to this work}
\author{Lillian~B.~Hughes$^2$}
\author{Nathaniel~T.~Leitao$^1$}
\author{Piotr~Put$^1$}
\author{Hengyun~Zhou$^{1,3}$}
\author{Nazli~U.~Koyluoglu$^{1,4}$}
\author{Simon~A.~Meynell$^5$}
\author{Ania~C.~Bleszynski~Jayich$^5$}
\author{Hongkun~Park$^{1,6}$}
\author{Mikhail~D.~Lukin$^1$}

\begin{abstract}

Electronic spins of nitrogen vacancy (NV) centers in diamond constitute a promising system for micro- and nano-scale magnetic sensing\cite{Schirhagl2014, Casola2018, Mamin2013, Hong2013}, due to their operation under ambient conditions\cite{Balasubramanian2009}, ease of placement in close proximity to sensing targets\cite{Ofori2012}, and biological compatibility\cite{Mohan2010}. At high densities, the electronic spins interact through dipolar coupling, which typically limits\cite{Zhou2020} but can also potentially enhance\cite{Kitagawa1993} sensing performance. Here we report the experimental demonstration of many-body signal amplification in a solid-state, room temperature quantum sensor. Our approach utilizes time-reversed two-axis-twisting interactions, engineered through dynamical control of the quantization axis and Floquet engineering\cite{Choi2020Floquet} in a two-dimensional ensemble of NV centers. Strikingly, we observe that the optimal amplification occurs when the backward evolution time equals twice the forward evolution time, in sharp contrast to the conventional Loschmidt echo\cite{PhysRevA.94.010102,Davis2016}. These observations can be understood as resulting from  an underlying time-reversed mirror symmetry of the microscopic dynamics, providing key insights into signal amplification and opening the door towards entanglement-enhanced practical quantum sensing.

\end{abstract}

\maketitle

Quantum correlations and entanglement are now actively explored 
to improve precision measurements in a variety of quantum systems.  
For example, in so-called 
spin squeezed states\cite{Kitagawa1993, Wineland1992}, quantum correlations are generated to reduce (squeeze) the quantum projection noise along a certain quadrature of the collective spin, thereby improving the precision of measurements along that quadrature.  Recently, spin squeezing has been realized in a variety of experimental platforms, including cavity-QED systems\cite{Pedrozo2020,Greve2022}, Rydberg atoms\cite{Bornet2023,Hines2023,Eckner2023}, trapped ions\cite{Franke2023}, Bose-Einstein condensates\cite{Muessel2014}, superconducting qubits\cite{Xu2020}, and atomic vapor cells\cite{Bao2020}, with broad applications including atomic clocks\cite{Pedrozo2020,Eckner2023}, magnetometry\cite{Muessel2014,Bao2020}, and matter-wave interferometry\cite{Greve2022}. While these approaches can be effective in systems with a high degree of isolation and control, 
an alternative strategy - signal amplification\cite{Davis2016} - focuses on using many-body dynamics for amplifying the measured signal, rather than reducing quantum noise
(Fig.~\ref{fig1}(a)).  While complementary to spin squeezing, 
the key feature  of this approach is the 
improved robustness against technical noise during readout, which makes it applicable to a much broader class of quantum systems 
that do not naturally support high fidelity quantum limited measurements under ambient conditions\cite{Wolf2015}.

Here we develop and demonstrate
a method for signal amplification in a two-dimensional (2D) ensemble of electronic spins associated with NV centers in diamond at room temperature. 
 Featuring long coherence times and well-developed quantum control techniques, such a system has recently emerged as a promising platform for micro- and nano-scale magnetic sensing applications\cite{Schirhagl2014, Casola2018, Mamin2013, Hong2013, Balasubramanian2009, Ofori2012, Mohan2010}. Despite the many applications of these solid-state quantum sensors, signal amplification has not yet been realized in such systems due to the short-ranged, anisotropic nature of the dipolar interaction and the faster decoherence compared to atomic platforms that are better isolated. Our approach overcome these challenges by a combination of dynamical control of the quantization axis to ameliorate the dipolar anisotropy, and Floquet engineering of two-axis-twisting\cite{Kitagawa1993} dynamics that exhibits fast amplification.
 Combining these techniques with time-reversed dynamics, we demonstrate a signal amplification of 6.7(6)\%.
 Remarkably, in contrast to prior approaches involving a symmetric time-reversed echo\cite{Colombo2022,LiZeyang2023},  we observe that the optimal amplification occurs under an asymmetric echo where the backward evolution time equals twice the forward evolution time. We show that these observations can be understood as resulting from the time-reversed mirror symmetry of the underlying Hamiltonian, which yields a generic, robust  mechanism for quantum enhancement in precision measurements. These results open the door towards new applications involving
entanglement-enhanced sensing in the solid state under ambient conditions.

\subsection*{Many-body  spin sensors in diamond}
Our experiments utilize a high density, dipolar interacting ensemble of electronic spins associated with NV centers in diamond. In order to enhance metrological sensitivity
through interactions, a common approach is to evolve the system under one-axis-twisting (OAT) or two-axis-twisting (TAT) Hamiltonians\cite{Kitagawa1993}. A unique challenge for dipolar interactions is their anisotropic nature, which causes spin pairs separated along and perpendicular to their quantization axis to interact with opposite signs (i.e. ferromagnetic versus anti-ferromagnetic coupling). This leads to zero net twisting when averaging over all orientations of spin pairs in three-dimensional spin ensembles. To overcome the dipolar anisotropy, we confine the NV centers to a 2D plane, achieved by nitrogen-$\delta$-doping\cite{Ohno2012,Hughes2023} during diamond chemical vapor deposition (CVD) growth (see Methods for details). The resulting sample consists of a positionally-disordered NV ensemble distributed within a 9-nm-thick (full-width half-max) layer, which is thinner than the typical NV-NV spacing ($\sim$17~nm), so that the geometry is approximately 2D.
\begin{figure}
\centering
\includegraphics[width=1.0\columnwidth]{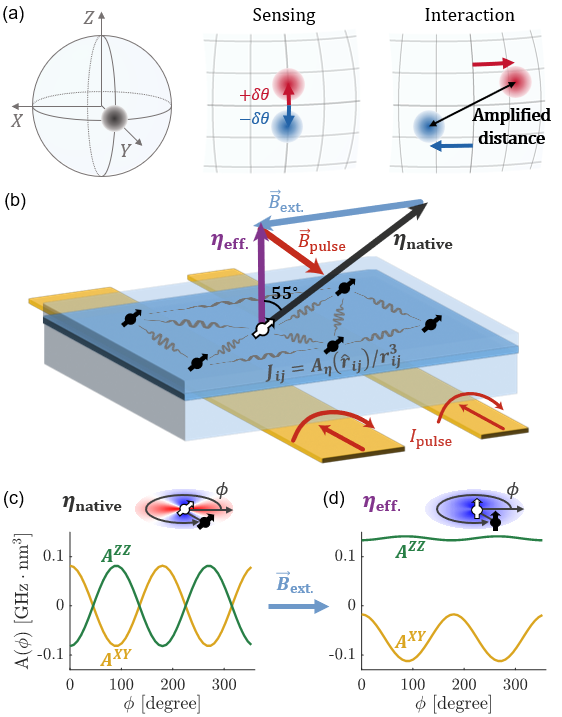}
\caption{\textbf{Engineering uniform-signed twisting interactions.} \textbf{a,} Concept of signal amplification. Different sensing signals lead to different states on the Bloch sphere, and the distance between them can be amplified under many-body interaction, resulting in enhanced sensitivity. The finite extent of the states represent technical noises during readout. The exact protocol with which we achieve signal amplification is shown in Fig.~\ref{fig4}(a). \textbf{b,} Schematic for the experimental system. The transparent box represents the diamond sample, with interacting NV centers confined in a two-dimensional plane within it. An external field (blue arrow) is used to change the quantization axis from the native axis ($\eta_{\mathrm{native}}$, black) to the normal vector of the plane ($\eta_{\mathrm{eff.}}$, purple). Additionally, a pulsed field (red, generated by the two wires below the diamond) is turned on during the initialization and readout, leading to a total field along the NV axis, which is required for high quality initialization and readout. \textbf{c,} Angular dependence of the Ising part (green) and exchange part (yellow) of the dipolar interaction for the native quantization axis. The color in the inset indicates the sign of $A^{ZZ}-A^{XY}$, which controls the OAT dynamics (see main text). \textbf{d,} Same as (c), but for the engineered quantization axis $\eta_{\mathrm{eff.}}$. The residual augular dependence is a consequence of the spin-1 nature of NV centers.
}
\label{fig1}
\end{figure}
The dipolar anisotropy can be mitigated if the quantization axis of the NV centers is perpendicular to the 2D plane, corresponding to (111) crystal axis orientation, as the separation of spin pairs would be predominantly perpendicular to the quantization axis. While  fabrication techniques for growing such 
(111)-oriented samples are under active development\cite{Hughes2024}, here we subject a (100)-oriented diamond to a strong ($\sim$890~G) magnetic field ${\vec{B}}_{\mathrm{ext}}$ (Fig.~\ref{fig1}(b), blue vector), changing the quantization axis from the native orientation ($\eta_{\mathrm{native}}$, black) to the plane normal ($\eta_{\mathrm{eff.}}$, purple). 

This configuration is motivated by the observation that the native quantization axis, tilted 55$^\circ$ from the plane normal, results in a mixed-signed and net zero dipolar interaction when averaged over the 2D plane (Fig.~\ref{fig1}(c)). While the engineered quantization axis ($\eta_{\mathrm{eff.}}$)
overcomes such problem (Fig.~\ref{fig1}(d)), the off-axis (i.e. not along the native NV axis) component of ${\vec{B}}_{\mathrm{ext}}$ mixes the $m=0$ and $m=\pm 1$ NV spin states, and thus disrupts the NV's native initialization and readout mechanisms\cite{Tetienne2012,Doherty2013}. Therefore, during initialization and readout, we cancel such off-axis field by turning on an auxiliary pulsed B field (Fig.~\ref{fig1}(b), red), which is adiabatically switched off in the middle to evolve the system under the twisting dynamics associated with the quantization axis $\eta_{\mathrm{eff.}}$. With the pulsed field, we improve the initial spin polarization by a factor of 4 (Extended Data Fig.~\ref{si_fig1}(d)), which yields a 4000-fold improvement in data averaging speed (see Supplement).

\subsection*{Engineering one- \& two-axis twisting dynamics}

To elucidate how OAT dynamics emerge from dipolar interactions, we decompose the XXZ Hamiltonian - derived from the secular approximation\cite{sakurai2020modern} (see Supplement) - into a Heisenberg term and a twisting term
\begin{align}
    \hat{H}_{\mathrm{XXZ}}
    =\sum_{i<j}\left[J_{ij}^{\mathrm{Heis}}{\hat{\vec{\sigma}}}_i\cdot{\hat{\vec{\sigma}}}_j+J_{ij}^{\mathrm{Twist}}\hat{\sigma}_i^z\hat{\sigma}_j^z\right],
\label{eq:Heisenberg_Ising_decomposition}
\end{align}
where $i$, $j$ index the spins (omitted from now on for simplicity), $\hat{\sigma}_i^{x,y,z}$ are the Pauli operators defined on the $m=0,-1$ electronic ground states of NV centers, and $J_{ij}^{\mathrm{Heis}}\equiv A_\eta^{XY}\left(\hat{r}_{ij}\right)/r_{ij}^3$, $J_{ij}^{\mathrm{Twist}}\equiv\left(A_\eta^{ZZ}\left(\hat{r}_{ij}\right)-A_\eta^{XY}\left(\hat{r}_{ij}\right)\right)/r_{ij}^3$ are defined with $A_\eta^{XY}$, $A_\eta^{ZZ}$ being the coefficients of the exchange and Ising parts of the dipolar coupling (Fig.~\ref{fig1}(c,d), see Supplement for calculations).
In this decomposition, the Heisenberg term does not affect the early-time dynamics of a spin-polarized initial state (which is an eigenstate of it), and the second term is a short-ranged OAT Hamiltonian. Therefore, as long as $\left(A^{ZZ}-A^{XY}\right)$ does not average to zero - as is the case for the interaction in Fig.~\ref{fig1}(d) - twisting dynamics can be obtained following an average over all spin pairs.

\begin{figure}
\centering
\includegraphics[width=1.0\columnwidth]{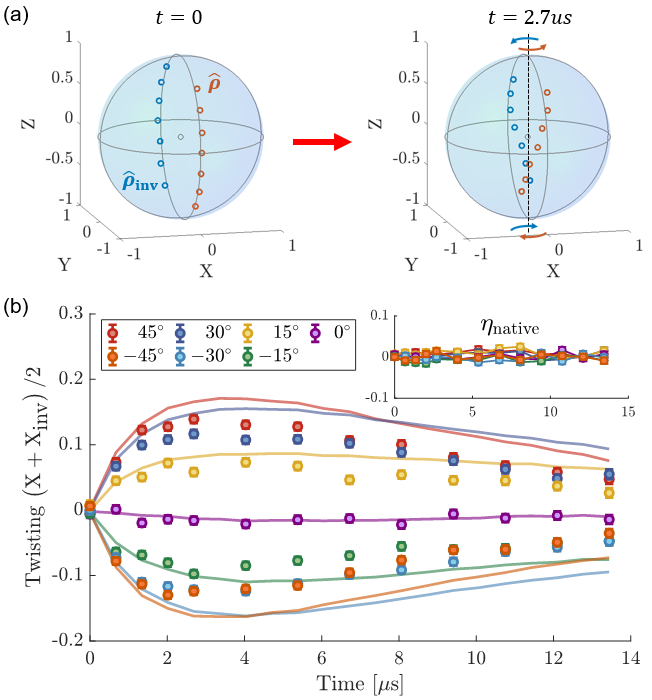}
\caption{\textbf{One-axis-twisting dynamics.}  \textbf{a,} Measured polarization dynamics of a series of initial states (left) in the YZ plane after 2.7~$\mathrm{\mu s}$ evolution (right). The red (blue) circles represent initial states on the front (back) side of the Bloch sphere. \textbf{b,} Twisting signal (in units of Bloch sphere radius) for the initial states in (a) with varying tilting angles from the equator to Z axis. The signal is averaged between antipodal pairs on the Bloch sphere ($\hat{\rho}$ and $\hat{\rho}_{\mathrm{inv}}$ in (a)) to cancel the effects from global rotations originating from pulse errors (Methods). The curves are predictions from our most realistic model of the system (Methods). Inset: same experiment with native quantization axis. Errors represent 1 s.d. accounting statistical uncertainties.
}
\label{fig2}
\end{figure}

To observe the OAT dynamics experimentally, we further suppress the on-site disorder with an XY8 pulse sequence\cite{Gullion1990}, and measure the dynamics of a series of initial states polarized in the YZ plane (Fig.~\ref{fig2}(a)). Here we see that initial states prepared in the upper hemisphere rotate counterclockwise, while initial states in the lower hemisphere rotate clockwise, consistent with the OAT dynamics.  
To further quantify the amount of twisting, the twisting signal, defined as the X polarization generated by the twisting dynamics, is measured on the same set of initial states (Fig.~\ref{fig2}(b)). In order to be robust against systematic global rotations originating from pulse errors (Methods, Extended Data Fig.~\ref{si_fig8}), the signal is averaged over pairs of antipodal states ($\hat{\rho}$ and $\hat{\rho}_{\mathrm{inv}}$ in Fig.~\ref{fig2}(a)). As expected, the data shows a $Z$-dependent twisting rate at early times, until eventually decaying after the interaction limited $T_2$ ($\sim$5~$\mu$s). To gain further evidence that the observed dynamics originate from dipolar interactions, we perform the same measurement at the native quantization axis $\eta_{\mathrm{native}}$ (Fig.~\ref{fig2}(b), inset), where no twisting dynamics are expected due to the dipolar anisotropy. The sharp contrast between them unambiguously rules out any alternative explanations based on systematic imperfections in the experiment.

\begin{figure}
\centering
\includegraphics[width=1.0\columnwidth]{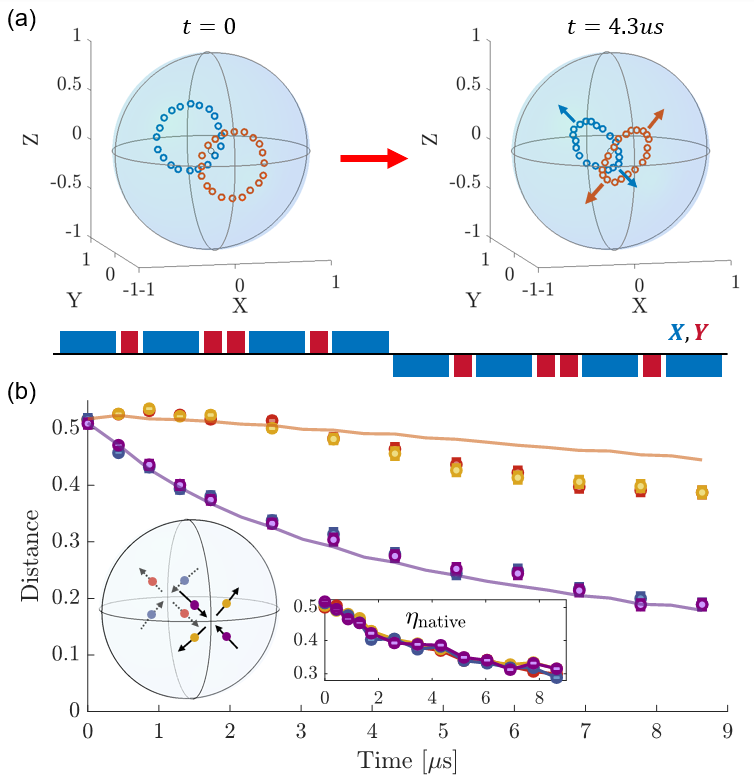}
\caption{\textbf{Two-axis-twisting dynamics.}  \textbf{a,} Measured polarization dynamics of a series of states that initially reside on two circles around $\pm\mathrm{Y}$ axis (left), after evolving for 4.3~$\mathrm{\mu s}$ under the TAT dynamics (right). Red (blue) circles represent initial states on the front (back) side of the Bloch sphere. The Floquet pulse sequence used to engineer the TAT Hamiltonian is plotted on the bottom, with upward blue (downward red) rectangles representing pulses along +X (-Y) axis. \textbf{b,} Measured distance (in units of Bloch sphere radius) between the initial state pairs with corresponding colors in the left inset, as a function of time evolved under the TAT dynamics. The initial states are tilted by 15° away from the $\pm\mathrm{Y}$ axis, leading to the initial distance $2\sin{\left(15^\circ\right)}$. The curves are predictions from our most realistic model of the system (Methods). Insets: Bloch sphere representation of the local semi-classical flow under the TAT dynamics, and the same experiment with native quantization axis. Errors represent 1 s.d. accounting statistical uncertainties.
}
\label{fig3}
\end{figure}

The resulting OAT dynamics can in principle be used for signal amplification in sensing experiments. This is quantified by the distance 
\begin{equation}
D\equiv \sqrt{\left(\Delta X\right)^2 + \left(\Delta Y\right)^2 + \left(\Delta Z\right)^2}
\label{eq:distance_definition}
\end{equation}
between two states that experienced opposite sensing rotations $\pm \delta\theta$ (Fig.~\ref{fig1}(a)), where $\Delta X,Y,Z$ are defined as the difference of the spin polarization between these two states. Under OAT dynamics, this distance is amplified for a pair of states initially separated along the Z direction (Fig.~\ref{fig1}(a)), as a non-zero $\Delta X$ is generated during the dynamics while the initial $\Delta Z$ is conserved. However, for the relatively weak twisting observed (Fig.~\ref{fig2}(b)), the amplification will be very small, as the increase of $D$ is second-order in $\Delta X$. Such small amplification can be easily overwhelmed by imperfect Z conservation coming from finite $T_1$ relaxation time and other imperfections. Therefore, dynamics that exhibit first-order amplification, such as TAT dynamics (see the flow diagram in Fig.~\ref{fig3}(b) inset), are more desirable.

The dipolar TAT Hamiltonian
\begin{equation}
\hat{H}_{\mathrm{TAT}}=J^{\mathrm{Heis}}\hat{\vec{\sigma}}\cdot\hat{\vec{\sigma}}+\lambda J^{\mathrm{Twist}}\left(\hat{\sigma}^z\hat{\sigma}^z-\hat{\sigma}^x\hat{\sigma}^x\right)
\label{eq:TAT_Hamiltonian}
\end{equation}
 can be engineered from the OAT Hamiltonian (Eq.~(\ref{eq:Heisenberg_Ising_decomposition}))
 by Floquet engineering\cite{Miller2024,Liu2011}. Motivated by the requirement to break the $U\left(1\right)$ symmetry and the good decoupling properties of XY16 sequence\cite{Gullion1990}, we designed a pulse sequence in which every $\pi$-pulse along the X direction is replaced by a $3\pi$-pulse (Fig.~\ref{fig3}(a), bottom), which explicitly breaks the symmetry between X-pulses and Y-pulses, while preserving the structures responsible for good decoupling properties. The exact TAT Hamiltonian, with the parameter $\lambda = \frac{2}{9}$, is then achieved by choosing appropriate values for the pulse durations and spacings (Methods). 

To confirm the engineering of TAT dynamics, we prepare a series of initial states residing on two circles around the $\pm\mathrm{Y}$ axis of the Bloch sphere, and measure their dynamics after a 4.3~$\mu$s evolution (Fig.~\ref{fig3}(a)). These states, which initially form two circles, deform into two ellipses with perpendicular major axes, consistent with the predicted semi-classical flow (Fig.~\ref{fig3}(b), left inset).  
Motivated by the goal of signal amplification, we further prepare initial states tilted by 15° from the $\pm\mathrm{Y}$ axis into the local amplifying and deamplifying directions (i.e. $X\pm Z$), depicted as the colored dots in Fig.~\ref{fig3}(b) left inset. We then measure the distance (Eq.~(\ref{eq:distance_definition})) between the two states in each pair with the same color, as a function of evolution time. As depicted in the left inset of Fig.~\ref{fig3}(b), the distances of the yellow and red pairs are expected to amplify, while the distances of the purple and blue pairs are expected to deamplify. In Fig.~\ref{fig3}(b), this shows up as a clear separation between the distances of the amplifying pairs and the distances of the deamplifying pairs, in sharp contrast to the no-separation case observed at the native quantization axis (right inset). The data show good agreement with theory, and 3.4(8)\% early-time amplification. 

\subsection*{Enhancing amplification through time-reversed echo}

To further enhance signal amplification, we note that  effects of interactions in the TAT Hamiltonian are two-fold: First, the mean-field TAT flow (Fig.~\ref{fig3}(b), left inset) leads to an increase of the distance (of the amplifying pairs). Second, the short-ranged nature of the dipolar interaction and the associated lack of total spin conservation also leads to rapid decay of the Bloch vector length\cite{Davis2023,Martin2023}, and thus a decrease of the distance. Motivated by the competition between these two mechanisms, we next explore if a time-reversed echo protocol could improve the amplification, as the decay can be (partially) reversed.

\begin{figure*}
\centering
\includegraphics[width=2.0\columnwidth]{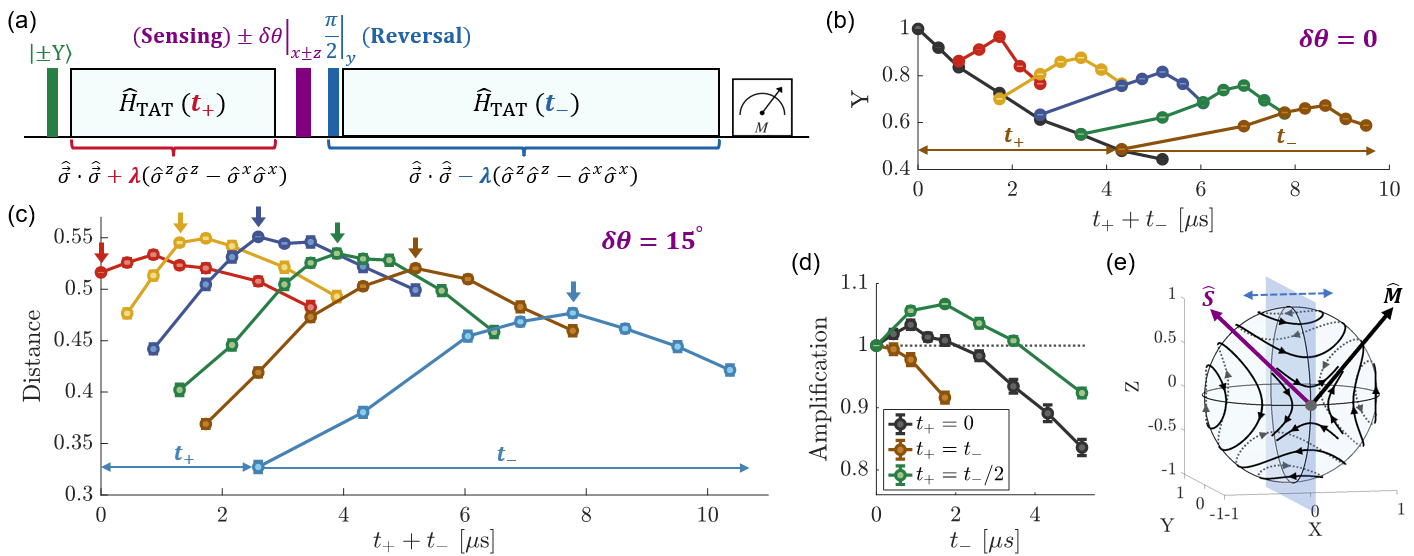}
\caption{\textbf{Time-reversed echo experiments.} \textbf{a,} The protocol of time-reversed echo, where a sensing rotation towards the local amplifying direction (purple pulse, simulated by explicit microwave driving) is sandwiched between forward (red, duration $t_+$) and backward (blue, duration $t_-$) evolution under the TAT Hamiltonian. The reversal is achieved by a $\frac{\pi}{2}$ pulse, which switches $\hat{\sigma}^x$ and $\hat{\sigma}^z$. \textbf{b,} Measurement of the Y revival under the time-reversal, without the sensing step ($\delta\theta=0$). The black curve is the reference decay without time-reversal, and each of the other curves corresponds to dynamics reversed at a different $t_+$, indicated by the leftmost point on the curve (e.g. for the brown curve, $t_+ = 4.3~\mu$s, as indicated by the brown horizontal arrows). \textbf{c,}  Signal amplification measurement (in units of Bloch sphere radius) sweeping both $t_+$ and $t_-$, with sensing angle $\delta\theta=15$°. Each curve represents a different $t_+$, similar to panel (b). The downward arrows indicate the locations where $t_-=2t_+$. The distance is averaged between the two amplifying pairs (see Fig.~\ref{fig3}(b)). \textbf{d,} Data in (c) replotted to compare the amplification without echo (black), with symmetric echo (brown), and with asymmetric echo (green). \textbf{e,} The optimal amplification at $t_+ = t_-/2$ can be explained by the time-reversed mirror symmetry of the TAT Hamiltonian, which reverses the flow direction and exchanges the sensing ($\hat{S}$) and measurement ($\hat{M}$) operators (see main text). Errors represent 1 s.d. accounting statistical uncertainties.}
\label{fig4}
\end{figure*}

Specifically, we use the echo protocol depicted in Fig.~\ref{fig4}(a), where a sensing rotation towards the local amplifying direction (simulated by explicit microwave driving) is sandwiched between a pair of time-reversed TAT Hamiltonians, engineered by Floquet sequences. While such echo protocols are typically implemented with equal duration of forward ($t_+$) and backward ($t_-$) evolutions\cite{Davis2016,Colombo2022,LiZeyang2023}, we explore  more general asymmetric echoes, where $t_+$ and $t_-$ are swept independently. To realize the time-reversed dynamics, we used a $\frac{\pi}{2}$-pulse to switch $\hat{\sigma}^x$ and $\hat{\sigma}^z$. Although the Heisenberg term $\hat{\vec{\sigma}}\cdot\hat{\vec{\sigma}}$ cannot be reversed, it has a trivial effect on 
early-time dynamics for spin-polarized initial states.

To verify the time-reversal protocol, we first measure the revival of the initial state polarized along the Y axis following the time-reversal, without the intermediate sensing step. As shown in Fig.~\ref{fig4}(b), the Y polarization decays under the TAT Hamiltonian during $t_+$, and then revives at $t_-=t_+$, confirming the validity of our implementation.
We then apply the echo protocol to signal amplification, measuring the distance between the two states in each amplifying pair (see Fig.~\ref{fig3}(b)) prepared by opposite sensing rotations $\pm \delta\theta$ (purple pulse in Fig.~\ref{fig4}(a)). The average distance between the two amplifying pairs is shown in Fig.~\ref{fig4}(c), where the leftmost point with $t_+=t_-=0$ corresponds to the non-interacting case, thus serving as a reference for measuring amplification. Here we observe a maximum amplification of 6.7(6)\% at $\left(t_+, t_-\right) = \left(0.864, 1.728\right)\mathrm{\mu s}$, and remarkably, for sufficiently long evolution times ($t_+> 0.5~\mu$s) we find that optimal signal amplification occurs at $t_-=2t_+$ (downward arrows in Fig.~\ref{fig4}(c)), in contrast to the conventional echo, where $t_-=t_+$.

\subsection*{Discussion and outlook}

We next turn to interpretations of the experimental results. We first note that 
the symmetric echo protocol is known to provide a Heisenberg-limited sensitivity and an $\mathcal{O}(\sqrt{N})$ amplification if the qubits in the system are all-to-all coupled\cite{Davis2016}. However, our observations clearly demonstrate that in 
 the disordered, short-range coupled system, it performs poorly 
 and in fact does not lead to any amplification (Fig.~\ref{fig4}(d)).   In contrast, the asymmetric echo with $t_+=t_-/2$, which was also recently predicted 
 theoretically  to  give a Heisenberg-limited sensitivity in the all-to-all interacting case\cite{Schulte2020}, clearly outperforms both the symmetric echo and non-echo protocols in our experiments (Fig.~\ref{fig4}(d)). 
To understand such superior performance of the asymmetric echo, we consider a linear response theory of the sensing protocol. Specifically, in a generic protocol where we sense a perturbation $\text{e}^{-i\left(\delta\theta\right) \hat{S}}$ by measuring the expectation value of an operator $\hat{M}$, the signal amplification is characterized by the dynamical susceptibility, as quantified by the Kubo formula\cite{Kubo1957}: 
\begin{align}
    \chi_{MS}(t_+,t_-) &\equiv \frac{\mathrm{d}{\langle\hat{M}}\rangle}{\mathrm{d}\left(\delta\theta\right)} =  -i\bra{\pm\mathrm{Y}} \lbrack \hat{M}(t_+-t_-), \hat{S}(t_+) \rbrack \ket{\pm\mathrm{Y}}
\label{eq:Kubo_formula}
\end{align}
involving the initial state polarized along $\pm\mathrm{Y}$ and the Heisenberg picture evolution of the sensing and measurement operators, happening at time $t_+$ and $t_+ - t_-$, respectively (see the time axis in Extended Data Fig.~\ref{si_fig3}(a)).
Using this formalism, the optimal condition $t_+=t_-/2$ can be understood as resulting from an underlying time-reversed mirror symmetry in the microscopic Hamiltonian. Specifically, we consider a symmetry transformation composed of a time-reversal (which transforms the Pauli vector $\hat{\vec{\sigma}}$ to $-\hat{\vec{\sigma}}$) followed by a $\pi$-rotation along the X axis. Together, these two steps flip the sign of $\hat{\sigma}^x$ while preserving $\hat{\sigma}^y$ and $\hat{\sigma}^z$, and thereby commute with the TAT Hamiltonian (Eq.~(\ref{eq:TAT_Hamiltonian})). This shows up in the semi-classical flow (Fig.~\ref{fig4}(e)) as a mirror symmetry that reverses the flow direction. Such symmetry, 
which exchanges the sensing and measurement operators (Fig.~\ref{fig4}(e)) and flips the sign of time, leads to 
 $ \chi_{MS}\left(t_+, t_-\right) =
  \chi_{MS}\left(t_- - t_+, t_-\right)$
(see Supplement for derivation). 
This relation predicts that the amplification is symmetric around $t_+ = t_-/2$, where the amplification is guaranteed to obtain an extreme value. 
 In the positionally-disordered spin ensemble governed by TAT Hamiltonian, we find that the naturally occurring strongly-coupled pairs of spins - the so-called ``spin dimers"\cite{Martin2023,Braemer2022,Franz2022} - dominate the dynamics and guarantee a maximum value at this location (see Supplement and Extended Data Fig.~\ref{si_fig3}). This conclusion is further confirmed by modifying the Hamiltonian such that  $t_+ = t_-/2$ corresponds to a local minimum instead of the maximum value (Extended Data Fig.~\ref{si_fig4}).
   In practice, decoherence breaks the exact amplification symmetry around $t_+ = t_-/2$
 and together with imperfect spin polarization, they reduce  
the amplification from 48\% expected in the ideal case to the observed value of 6.7\% (Extended Data Fig.~\ref{si_fig6}).

Our results can be extended along several directions. 
A natural goal is to realize larger, scalable (i.e. improve with system size $N$) amplification, to achieve practical benefits in the solid-state sensing platform. Our numerical analysis shows that by reducing the positional disorder towards a 2D lattice\cite{Chen:19}, suppressing the dynamical on-site disorder by going to lower temperature\cite{Takahashi2008},  and improving the spin polarization\cite{Lee2024},
an improvement of the amplification above 10 (i.e. 20~dB) is expected theoretically (Extended Data Fig.~\ref{si_fig6}). 
Furthermore, a promising route towards scalable amplification would be to engineer interactions that simultaneously support XY ferromagnetic ordering\cite{Block2023}, which can be realized by advances on the growth of (111)-oriented diamonds\cite{Hughes2024} and reduction of positional disorder\cite{Kwasigroch2017} (Supplement, Extended Data Fig.~\ref{si_fig11}). In addition, further improvements of the practical sensitivity can be obtained by increasing the spin readout fidelity\cite{Shields2015,Arunkumar2023}. Combining our results with such methods could enable exploring new opportunities beyond signal amplification, since quantum projection noise would become relevant in such cases. Ultimately, combined with other techniques developed for solid-state quantum sensors, such as the capability of bringing the sensors to close proximity with sensing targets\cite{Ofori2012} and the injection of sensor-containing nanoparticles for in vivo measurements\cite{Mohan2010,Choi2020Worm}, the present approach can lead to practical advances in nanoscale imaging technologies for novel condensed matter materials and biological structures.

Beyond the direct applications to dense NV ensembles, the present observations demonstrate a generic and robust mechanism for exploiting metrologically useful entanglement, applicable to a wide variety of sensing platforms. In particular, the 
time-reversed mirror symmetry applies to a vast majority of Hamiltonians used for entanglement-enhanced sensing, including the OAT Hamiltonian, TAT Hamiltonian, LMG Hamiltonian\cite{LiZeyang2023}, and even a cubic XYZ Hamiltonian\cite{Zhang2024}. Although such symmetry only guarantees an extreme value of amplification at $t_+ = t_-/2$ instead of a maximum value, we show that a maximum is expected in most cases (Supplement, Extended Data Fig.~\ref{si_fig2}), suggesting the generality of the observed optimal amplification under asymmetric echo. Indeed, as discussed in Ref.\cite{Schulte2020} and our complementary work\cite{AmpTheory}, asymmetric echo with $t_+ = t_-/2$ can be exploited in systems exhibiting collective dynamics, theoretically enabling Heisenberg-limited sensing and an $\mathcal{O}(N)$ amplification, in platforms ranging from polar molecules\cite{Bilitewski2021} to trapped ions\cite{Britton2012} and cavity-QED systems\cite{Colombo2022}.

\bibliography{main.bib}

\clearpage
\newpage

\section*{Methods}

\noindent\textbf{Diamond sample} \\
Diamond homoepitaxial growth and delta doping\cite{Hughes2023} were performed via plasma-enhanced chemical vapor deposition (PECVD) using a SEKI SDS6300 reactor on a (100)-oriented electronic grade diamond substrate (Element Six Ltd.). Prior to growth, the substrate was fine-polished by Syntek Ltd. to a surface roughness of $\sim$200-300~pm, followed by a $4-5~\mathrm{\mu}$m etch to relieve polishing-induced strain. The growth conditions consisted of a 750~W plasma containing 0.1$\%$ $^{12}$CH$_{4}$ in 400~sccm H$_2$ flow held at 25~torr and $\sim$790~$^{\circ}$C according to a pyrometer. A $\sim$420~nm-thick isotopically purified (99.998$\%$ $^{12}$C) epilayer was grown. During the nitrogen $\delta$-doping period of growth, $^{15}$N$_2$ gas (1.0$\%$ of the total gas content) is introduced into the chamber for 30 minutes. After growth, the sample was characterized with secondary ion mass spectrometry (SIMS) to estimate the isotopic purity, epilayer thickness, and the thickness of the $\delta$-doped layer (8-10~nm full-width half-max).

The diamond was further electron irradiated and annealed to generate enhanced NV center concentrations. Irradiation was performed with the 200~keV electrons of a transmission electron microscope (TEM, ThermoFisher Talos F200X G2 TEM). The irradiation time was varied to create spots that range in dose from $10^{17}$-$10^{21}$~e$^{-}$/cm$^{2}$, and the presented experiments are performed at a spot with a dose of $2.8\times10^{20}$~e$^{-}$/cm$^{2}$. The sample then underwent subsequent annealing at 850~$^{\circ}$C for 6 hours in an Ar/H$_2$ atmosphere, during which the vacancies diffuse and form  NV centers. After irradiation and annealing, the sample was cleaned in a boiling triacid solution (1:1:1 H$_2$SO$_4$:HNO$_3$:HClO$_4$) and annealed in air at 450~$^{\circ}$C to oxygen terminate the surface and help stabilize the negative NV$^{-}$ charge state for further measurements. \\

\noindent\textbf{Circuit designs for pulsed current}\\
The pulsed current is controlled by a custom voltage controlled current source (Extended Data Fig.~\ref{si_fig1}(b)), with a 3~$\mathrm{\mu s}$ rising time and a 7~$\mathrm{\mu s}$ falling time. These times are much slower than the qubit frequency (712.24~MHz), but much faster than the $T_1$ time (0.94~ms), ensuring adiabatic switching of quantization axis while avoiding significant polarization loss during the process. The pulsed current is combined with the microwave driving using a bias-tee (Extended Data Fig.~\ref{si_fig1}(a)), and then delivered to the diamond through a coplanar waveguide (CPW). At the center of the CPW, the current flows through two parallel wires (Extended Data Fig.~\ref{fig1}(c)), and the presented experiments are done at a confocal spot between them. The thicknesses of these two wires are 10~$\mathrm{\mu m}$ and 4~$\mathrm{\mu m}$, and the gap between them is 6~$\mathrm{\mu m}$. These dimensions are optimized for the direction and strength of the pulsed field, and the homogeneity of the Rabi drive.

In the experiment, a pulsed field with strength 168~G and direction {23}$^\circ$ below the horizontal plane is generated by a 1~A current, resulting in a qubit frequency $f_{\mathrm{qubit}}$=712.24~MHz during the twisting dynamics and $f_{\mathrm{read}}$=660~MHz during initialization and readout. Operating at $\sim$25\% duty cycle, the pulsed current leads to significant heating. To prevent thermal failure of the coil, the circuit was fabricated on sapphire substrate with a heat conductivity one order of magnitude better than glass. With the help of the pulsed field, the spin polarization is improved by a factor of 4 (Extended Data Fig.~\ref{si_fig1}(d)).
\\

\noindent\textbf{Floquet sequences timings} \\
The timings in the TAT (and time-reversed TAT) engineering pulse sequence (Fig.~\ref{fig3}(a), bottom) are chosen as $t_\pi=12$~ns and $\tau=3$~ns, where $t_\pi$ and $\tau$ are $\pi$-pulse duration and pulse spacing, respectively. This choice is made through the following considerations: Firstly, to ensure the generation of TAT Hamiltonian, the effective time spent in the X, Y, and Z frames must satisfy $t_x+t_z=2t_y$\cite{Choi2020Floquet}, which dictates the ratio between $t_\pi$ and $\tau$. Secondly, it is desirable to repeat the sequence as fast as possible, to decouple the high frequency components of on-site disorder and thereby improve the coherence properties. Thirdly, the off-axis magnetic field used in this work leads to complicated nuclear spin dynamics (see Supplement), where accidental entanglement between the NV electronic spin and nuclear spin could happen when the Floquet period matches the nuclear spin precession period, which happens around  $t_\pi=24$~ns (Extended Data Fig.~\ref{si_fig7}). The choice of pulse timings needs to avoid these special durations to effectively decouple the nuclear spins. Finally, to prevent neighboring pulses from overlapping due to finite rising and falling time, $\tau$ cannot be smaller than 3~ns experimentally. The above considerations together lead to the timing choice in the experiment. The timings in the OAT dynamics are chosen arbitrarily as $t_\pi=38$~ns and $\tau=46$~ns, and did not require extensive optimization.
\\

\noindent\textbf{Controlling systematic errors} \\
Due to the relatively small amplification observed in this work, it is crucial to understand and control potential systematic errors, to make trustworthy interpretations of the experimental data. First, slow drift of the laser power causes drift of fluorescence (FL) intensity. This is overcome by differential readout, where we pair up each measurement with a variant of itself with an additional $\pi$-pulse before the readout, and focus on the contrast (i.e. fractional difference of the FL) instead of the raw FL values. This significantly reduces the effects of FL drift, but we still observe residual drift of the full contrast (i.e. contrast between +Z and -Z states), coming from different saturation levels of the single photon counting module (SPCM) under different FL intensity. This residual effect is further normalized out by interleaving the main experiments with measurements of the full contrast, on a time scale of 4 ms, which is much faster than the drift timescale. Similarly, we also interleave the main experiments with measurements of qubit frequency and Rabi frequency, and use feedback to stabilize them. The qubit frequency is stabilized to within $\pm$0.3~MHz, which is much smaller than the on-site disorder; and the Rabi frequency is stabilized to within $\pm$0.5\%.

In addition to the drift effects mentioned above, an important source of systematic error in the extraction of OAT signal (Fig.~\ref{fig2}(b)) is the global rotation of the Bloch sphere, coming from accumulated coherent pulse errors. As illustrated in Extended Data Fig.~\ref{si_fig8}(a), such rotations could locally mimic the twisting dynamics. This is overcome by averaging the measured X polarization on antipodal pairs on the Bloch sphere, since antipodal pairs remain antipodal under global rotations.

Although the distance measurements (Fig.~\ref{fig3},\ref{fig4}) are naturally insensitive to global rotations, problems can arise if the rotations are too large. As an extreme example, if a net $\frac{\pi}{2}$ rotation angle is accumulated along the Y axis, the amplifying pairs will transform into the deamplifying pairs and thereby affects the efficiency of amplification. Moreover, distortions of the pulse envelope along the microwave circuit lead to (small) overlaps of the sensing pulse with surrounding Floquet pulses, which could affect the effective sensing angle through interference. To control the pulse distortions, we measured the S parameter of the microwave circuit in the Fourier domain with a vector network analyzer (assuming the distortions being predominantly linear), and pre-distorted the waveform sent to the arbitrary waveform generator (AWG) with the inverse transformation to cancel the distortions. The effects of pulse pre-distortion are shown in Extended Data Fig.~\ref{si_fig9}. To be extra cautious, we further cancel the effects of the residual (tiny) pulse overlaps by averaging the measured distance between the two experiments shown in Extended Data Fig.~\ref{si_fig10}, where the overlap between the sensing pulse and surrounding pulses are negated, while the engineered Hamiltonian is kept the same.\\

\noindent\textbf{Numerical models}\\
All numerical simulations in this work are done using the cluster discrete truncated Wigner approximation (cluster-DTWA) method\cite{Wurtz2018,Braemer2024}, with each cluster including two spins. The clustering strategy we use is the same as Ref.\cite{Braemer2024}, which has been shown to give high accuracy results in strongly disordered systems. In our most realistic model, which was used to generate the theory predictions in Fig.~\ref{fig2}(b) and Fig.~\ref{fig3}(b), we considered the imperfections coming from finite 2D layer thickness, static on-site disorder, dynamical on-site disorder, and imperfect spin polarization, in addition to the ideal Hamiltonian evolution. Here, the 2D layer thickness (9~nm, FWHM) is measured with secondary ion mass spectroscopy (SIMS), the static on-site disorder (1.0~MHz, FWHM) is characterized through Ramsey measurements, the dynamical on-site disorder ($0.019~\mathrm{MHz}/\sqrt{\mathrm{MHz}}$ around the filter function peak $f_{\mathrm{peak}}=37~\mathrm{MHz}$) is characterized through spin-locking measurements\cite{Choi2017}, and the spin polarization (75\%) is estimated based on the NV rate equation\cite{Song2020} prediction (85\%) and measured contrast loss (10\%) during the quantization axis switching. The effects of these imperfections on amplification are studied numerically by turning them off one-by-one in the simulations, as shown in Extended Data Fig.~\ref{si_fig6}. All reported simulations in this work have system size $N=200$.\\

\noindent\textbf{Data Availability}\\
The data that supports the findings of this study are available from the corresponding author on request.\\

\noindent\textbf{Acknowledgements}
We thank  Emily J. Davis, Weijie Wu, Bingtian Ye, Zilin Wang, Norman Y. Yao, Szymon Pustelny, Nishad Maskara, Mathew Mammen, Siddharth Dandavate, Andrew Maccabe for helpful discussions and James MacArthur for technical contributions. This work was supported
by  the National Science Foundation (grant number PHY-2012023), the Center for Ultracold Atoms (an NSF Physics Frontiers Center), Gordon and Betty Moore Foundation Grant No. 7797-01, the U.S. Department of Energy [DOE Quantum Systems Accelerator Center (Contract No.: DE-AC02-05CH11231) and BES grant No. DE-SC0019241], and the Army Research Office through the MURI program grant number W911NF-20-1-0136. We acknowledge the use of shared facilities of the UCSB Quantum Foundry through Q-AMASE-i program (NSF DMR-1906325), the UCSB MRSEC (NSF DMR 1720256), and the Quantum Structures Facility within the UCSB California NanoSystems Institute. 
A.B.J. acknowledge support from the NSF QLCI program through grant number OMA-2016245. 
L.B.H. acknowledges support from the NSF Graduate Research Fellowship Program (DGE 2139319) and the UCSB Quantum Foundry.
\\

\noindent\textbf{Author contributions} H.G. and L.S.M. performed the experiments and analyzed the data. L.B.H. and S.A.M. fabricated the diamond sample. H.G. and N.T.L. developed theoretical interpretation of the asymmetric echo response. P.P., H.Z., and N.U.K. contributed ideas to the project. A.C.B.J., H.P., and M.D.L. supervised the project. All authors discussed the results and contributed to the manuscript.
\\

\noindent\textbf{Competing interests:} The authors declare no competing interests.\\ 

\noindent\textbf{Correspondence and requests for materials} should be addressed to M.D.L.\\

\setcounter{figure}{0}
\newcounter{EDfig}
\renewcommand{\figurename}{Extended Data Fig.}

\begin{figure*}
\centering
\includegraphics[width=2.0\columnwidth]{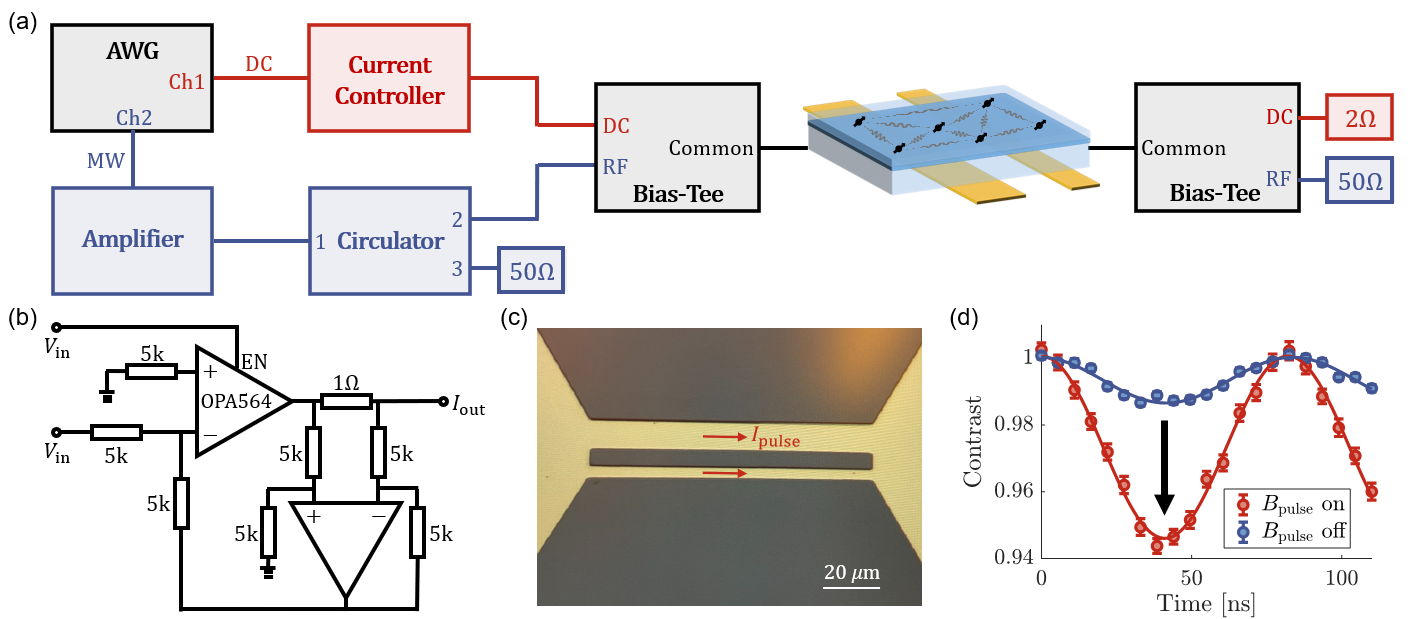}
\caption{\textbf{The circuit design.} \textbf{a,} Schematic of the whole circuit used in this work. In this diagram, the microwave components are labelled in blue, the pulsed current components are labelled in red, and components common to both are labelled in gray. The arbitrary waveform generator (AWG) is Tektronix AWG7122C, the amplifier is Mini-Circuits ZHL-1000-3W, the circulator is DBwave PACL1700600085A, and the two bias-tees are both UMCC BT-H250-LN. The 2~$\Omega$ resistor at the end of the circuit is used for monitoring the pulsed current. \textbf{b,} Circuit diagram of our custom current controller. \textbf{c,} Microscope image of the central feature of the coplanar waveguide. The thicknesses of the two wires are 10$~\mu$m and 4$~\mu$m, and the separation between them is 6$~\mu$m. In this work, we work at a confocal spot between the two wires. \textbf{d,} Rabi oscillation measurements with the pulsed field turned on during both initialization and readout (red), versus turned on only during readout (blue). A factor of 4 improvement of the contrast is observed, indicating a factor of 4 improvement of the initial spin polarization. Errors represent 1 s.d. accounting statistical uncertainties.
}
\refstepcounter{EDfig}\label{si_fig1}
\end{figure*}

\begin{figure*}
\centering
\includegraphics[width=2.0\columnwidth]{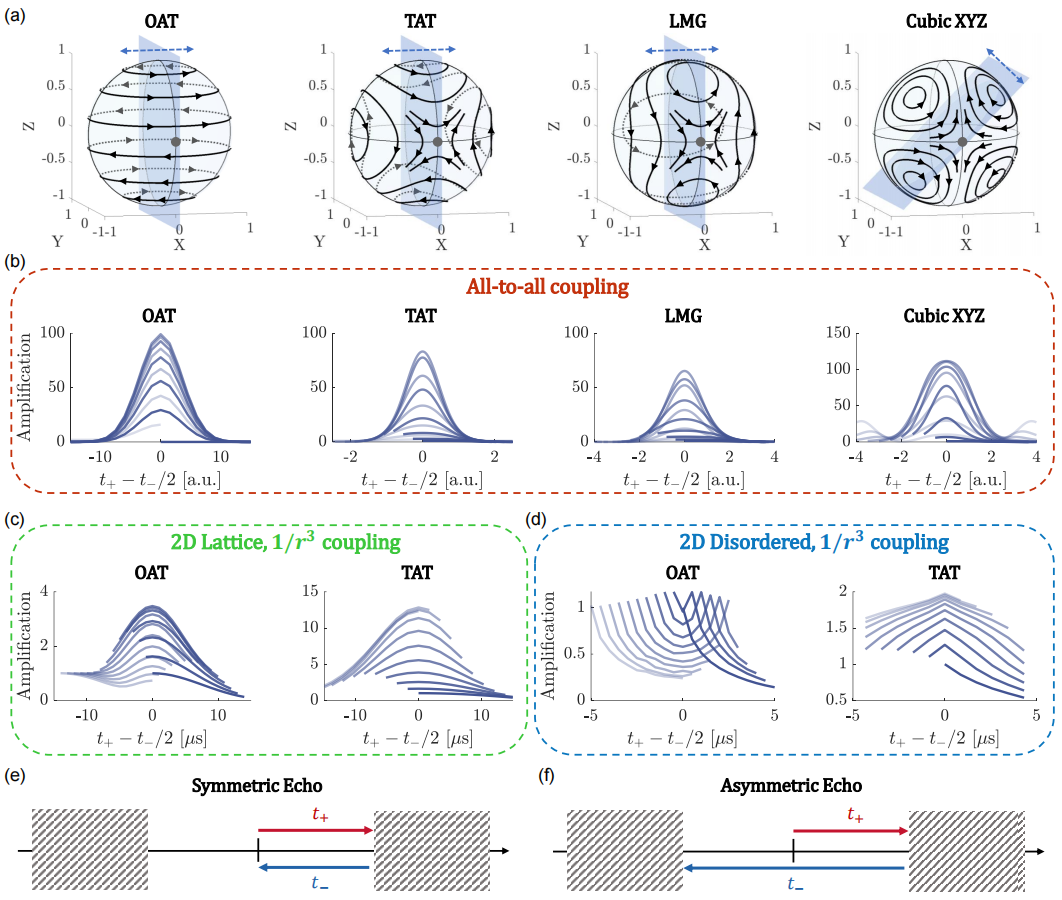}
\caption{\textbf{Generality of asymmetric echo.}  \textbf{a,} Time-reversed mirror symmetry of the considered Hamiltonians (see Supplement for their explicit expressions), where the flow directions are reversed under the mirror reflections. The gray dots at -Y axis indicates the initial states, which are preserved under the symmetries. \textbf{b,} Simulated amplification for these Hamiltonians, under the assumption that the coupling is all-to-all. In these plots, each trace indicates a different $t_-$, and darker traces corresponds to smaller $t_-$ values. \textbf{c, d,} Same as (b), but for $1/r^3$ coupling on a 2D lattice and positionally-disordered 2D ensemble. \textbf{e, f,} Heuristic argument for why $t_+=t_-/2$ gives a maximum amplification in most cases: Asymmetric echo allows doubling the amplifying time $t_-$ without encountering the highly over-twisted regime (shaded zones). Note that this heuristic argument is not a rigorous statement, as amplification is not necessarily degraded in the highly over-twisted regime (see our complementary work\cite{AmpTheory}).
}
\refstepcounter{EDfig}\label{si_fig2}
\end{figure*}

\begin{figure*}
\centering
\includegraphics[width=1.08\columnwidth]{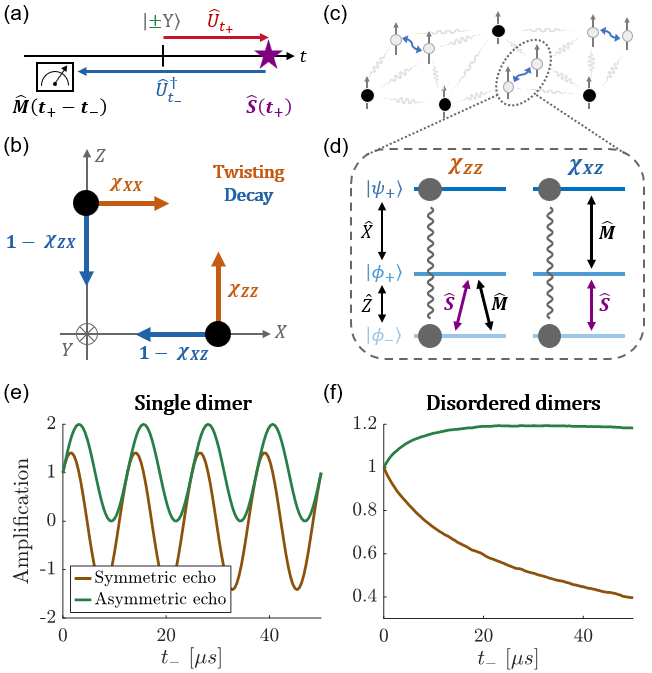}
\caption{\textbf{Microscopic mechanism of asymmetric echo response.} \textbf{a,} Heisenberg picture illustration of the asymmetric echo, where the sensing (purple star) and measurement (black box) operators happens at time $t_+$ and $t_+ - t_-$, respectively. Their anti-matched phase accumulation at $t_+ = t_-/2$ explains the observed maximum amplification at such time (see Supplement and our complementary work\cite{AmpTheory}). \textbf{b,} Physical interpretation of the susceptibility matrix. Black dots are the resulting states after sensing along X and Z, for non-interacting sensors. Orange (blue) arrows represent further twisting (decay) dynamics, as described by the diagonal (off-diagonal) components of the susceptibility matrix. \textbf{c,} Spin dimers (gray spins with blue arrow between them) naturally occur in a positionally-disordered spin ensemble. \textbf{d,} Energy spectrum of an isolated spin dimer, and the ``$\Lambda$" (``ladder") type  processes that describe the diagonal (off-diagonal) components of the susceptibility matrix. The gray disks and the curvy lines connecting them represent the initial state $\ket{\pm\mathrm{Y}}$, which is a coherent superposition $\ket{\phi_-} \pm i\ket{\psi_+}$. \textbf{e,} Analytically calculated amplification of an isolated dimer coupled by TAT Hamiltonian with coupling strength $J_{\mathrm{D}} = 2\pi\times 40~$kHz, under the symmetric echo (brown) and asymmetric echo with $t_+ = t_-/2$ (green). \textbf{f,} Same as (e), after averaging over positional disorder.
}
\refstepcounter{EDfig}\label{si_fig3}
\end{figure*}

\begin{figure*}
\centering
\includegraphics[width=0.54\textwidth]{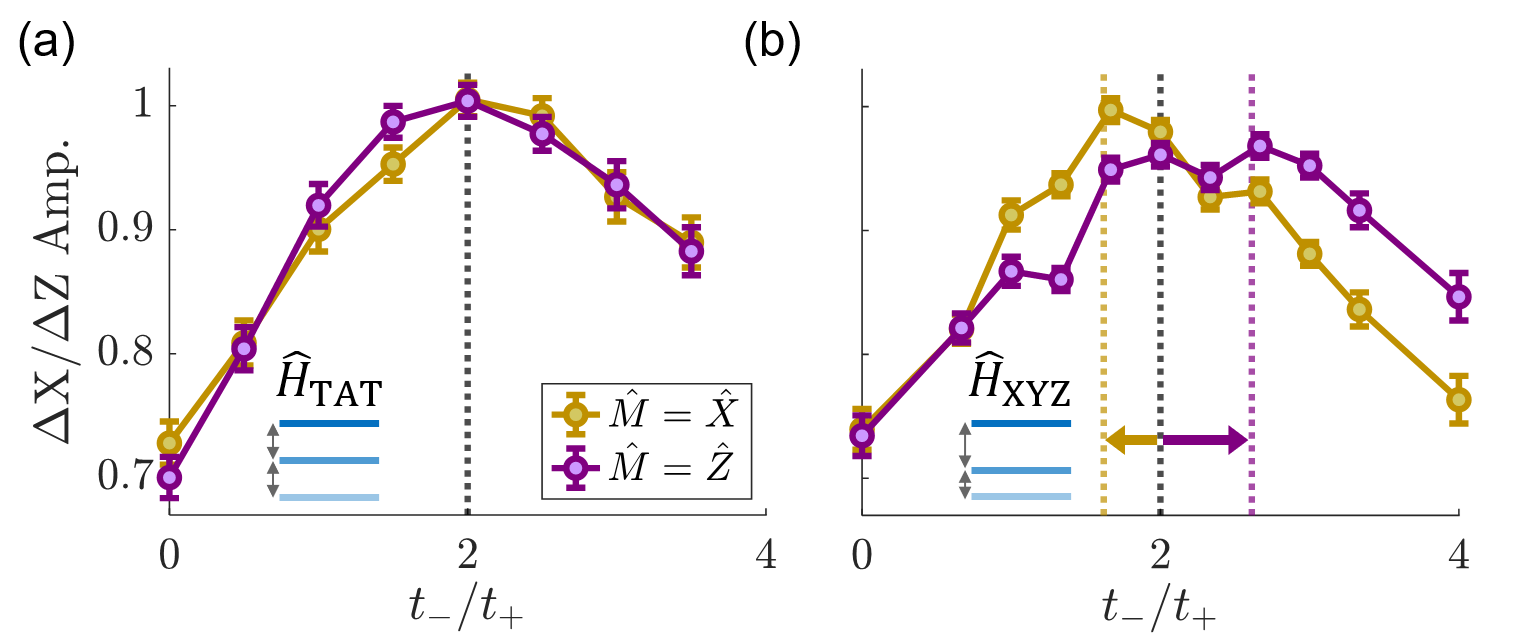}
\caption{\textbf{Further experimental evidence for the dimer mechanism.} \textbf{a,} Plotting the amplification of $\Delta X$ and $\Delta Z$ separately for the dataset in Fig.~\ref{fig4}(c) of main text, focusing on $t_+=1.73~\mathrm{\mu s}$ case. The harmonic dimer spectra (inset) explains the common peaks at $t_-=2t_+$. \textbf{b,} Same as (a) but measured under an XYZ Hamiltonian (see Supplement) that alters the dimer spectra (inset). The anharmonic spectra leads to a splitting of the peaks. The two colored dashed vertical lines represent the predicted peak locations based on the dimer spectra. Errors represent 1 s.d. accounting statistical uncertainties.
}
\refstepcounter{EDfig}\label{si_fig4}
\end{figure*}

\begin{figure*}
\centering
\includegraphics[width=1.0\textwidth]{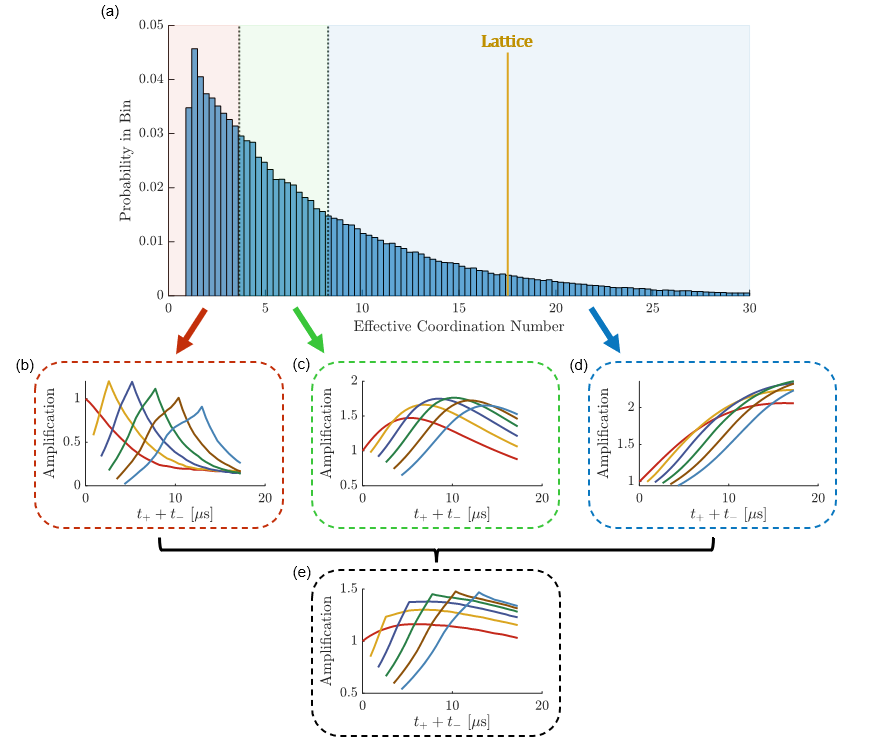}
\caption{\textbf{Contributions from spins with varying effective coordination number.} \textbf{a,} Histogram of the effective coordination number, as defined in Eq.~(\ref{eq:definition_effective_coordination_number}). The histogram is based on 1000 randomly sampled positional configurations of 200 spins (i.e. $2\times 10^5$ spins in total). The red, green, and blue shadings represent the groups of spins with low, middle, and high coordination number, each consisting $1/3$ of the total spin number. The yellow vertical line indicates the effective coordination number of a square lattice, as a reference. \textbf{b-d,} Average amplification for spins within each group, simulated without considering experimental imperfections. The plots have the same style as Fig.~\ref{fig4}(c) of main text, where each trace indicates a different $t_+$. \textbf{e,} The amplification of the full dipolar system can be reconstructed by averaging over (b-d).
}
\refstepcounter{EDfig}\label{si_fig5}
\end{figure*}

\begin{figure*}
\centering
\includegraphics[width=2.0\columnwidth]{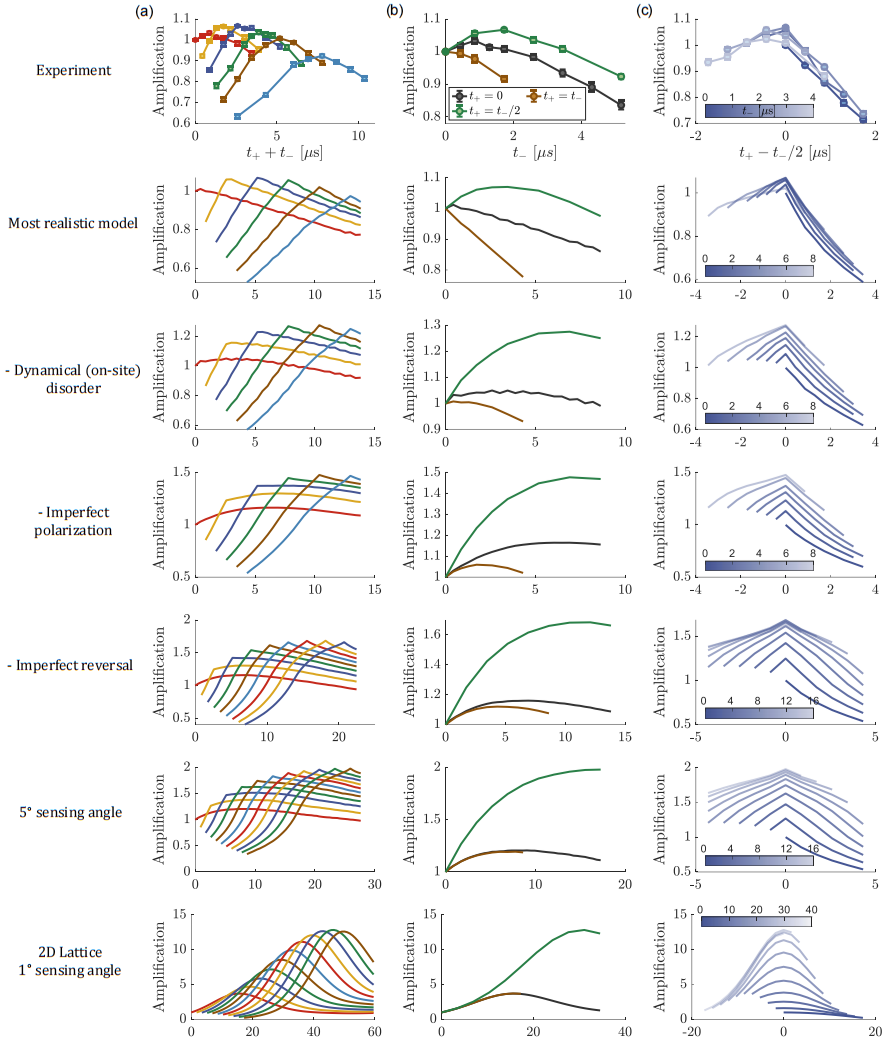}
\caption{\textbf{Understanding the experimental imperfections.} The asymmetric echo of TAT dynamics is simulated under different assumptions (rows), starting from the experimental data and gradually removing various experimental imperfections (see Supplement for discussions). The simulated amplification is plotted in three ways (columns): \textbf{a,} The same style as Fig.~\ref{fig4}(c) of main text, where each trace indicates a different $t_+$. \textbf{b,} The same style as Fig.~\ref{fig4}(d), comparing asymmetric echo to symmetric echo and non-echo. \textbf{c,} Visualizing the predicted symmetry around $t_+ = t_-/2$, where each trace indicates a different $t_-$, similar to Extended Data Fig.~\ref{si_fig2}. All simulations are done under the experimental value of the sensing angle $\delta\theta=15^\circ$, except the last two rows, where the sensing angles are reduced to probe the linear response. Errors represent 1 s.d. accounting statistical uncertainties.
}
\refstepcounter{EDfig}\label{si_fig6}
\end{figure*}

\begin{figure*}
\centering
\includegraphics[width=1.08\columnwidth]{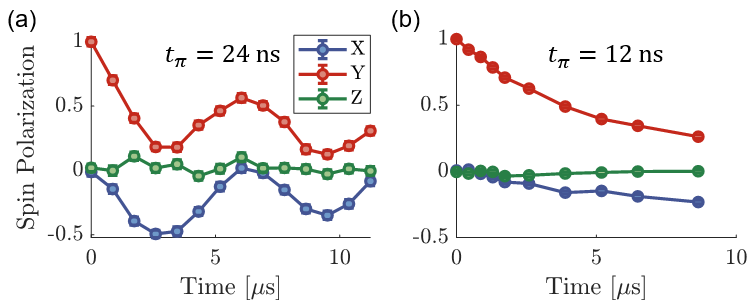}
\caption{\textbf{Nuclear spin decoupling.} \textbf{a,} Accidental entanglement between the NV electronic spin and nuclear spin happens when the Floquet period matches the nuclear spin precession period (see Supplement), showing up as oscillations in the measured spin polarization that cannot be explained by a global rotation of the Bloch sphere. The initial state in this experiment is a spin-polarized state along Y. \textbf{b,} These oscillations disappear when the Floquet period does not match the nuclear spin precession period. The (weak) rotation from +Y to -X comes from accumulated coherent pulse errors. Errors represent 1 s.d. accounting statistical uncertainties.
}
\refstepcounter{EDfig}\label{si_fig7}
\end{figure*}

\begin{figure*}
\centering
\includegraphics[width=1.8\columnwidth]{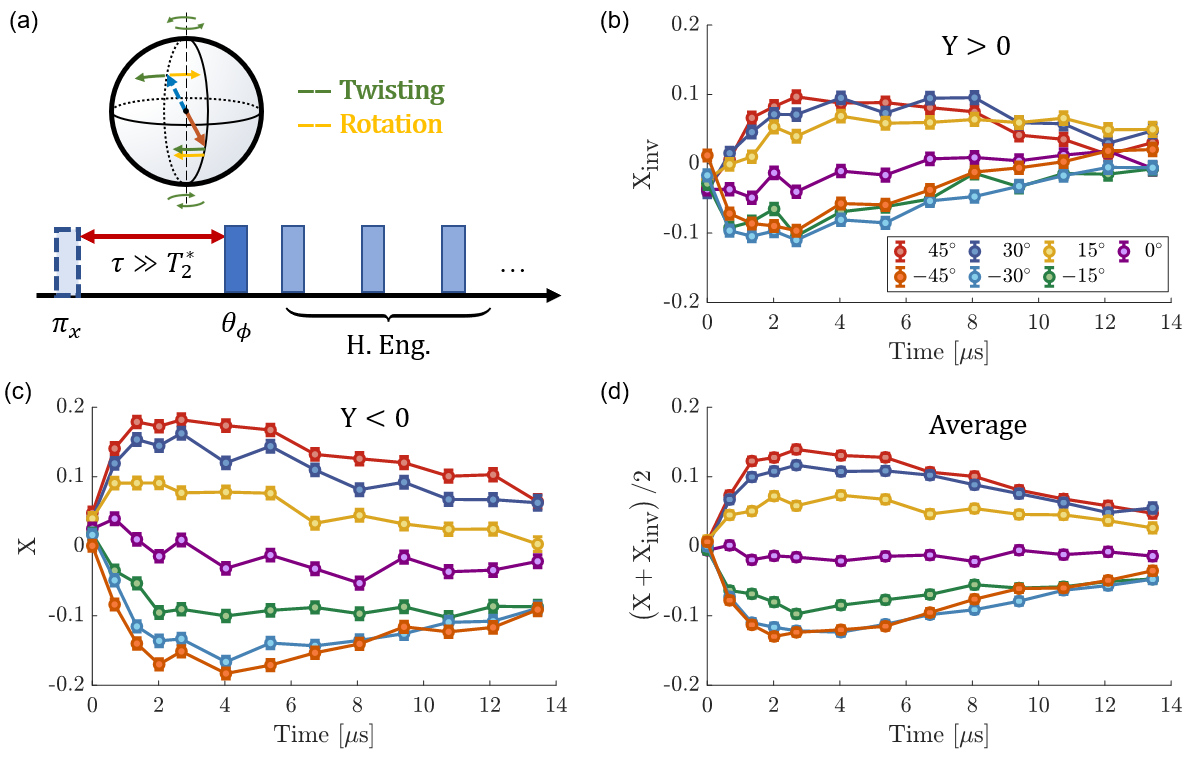}
\caption{\textbf{Robust OAT signal extraction.} \textbf{a,} Top: global rotations can mimic the twisting dynamics locally, but they can be distinguished by averaging between antipodal pairs. Bottom: utilizing dephasing to robustly prepare antipodal pairs (i.e. with and without the dashed $\pi$-pulse). \textbf{b, c,} Measured X polarization dynamics on the back (front) side of the Bloch sphere. \textbf{d,} Averaging (b) and (c) leads to the data reported in Fig.~\ref{fig2}(b), repeated here for convenience. Errors represent 1 s.d. accounting statistical uncertainties.}
\refstepcounter{EDfig}\label{si_fig8}
\end{figure*}

\begin{figure*}
\centering
\includegraphics[width=1.08\columnwidth]{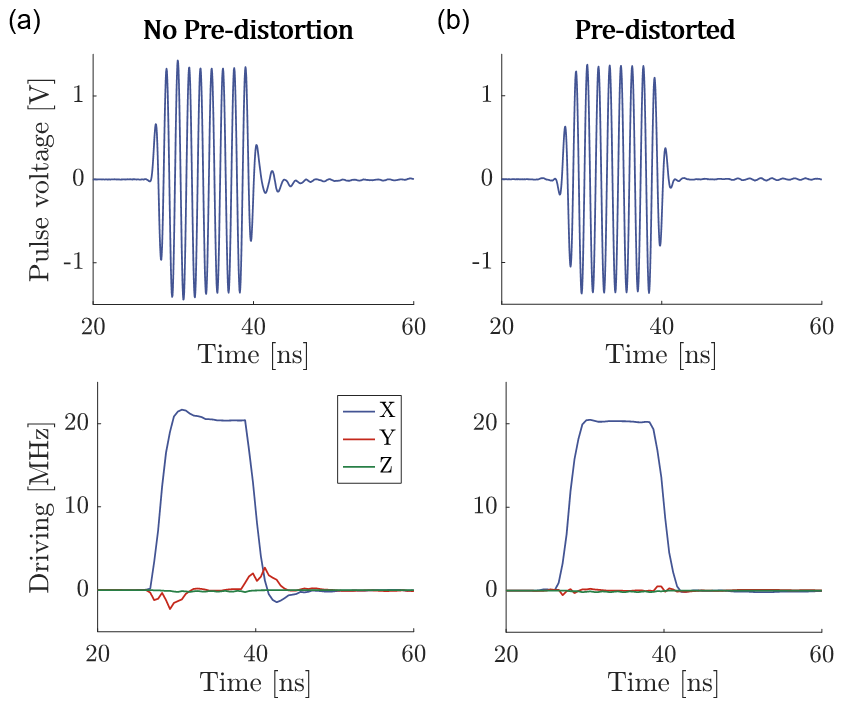}
\caption{\textbf{Effects of pulse pre-distortion.} \textbf{a,} A 12~ns duration $\frac{\pi}{2}$-pulse measured at the input of the coplanar waveguide (Extended Data Fig.~\ref{si_fig1}(a)), without implementing pulse pre-distortion. Top: raw waveform recorded on the oscilloscope. Bottom: Rabi frequency in the rotating frame, after taking the rotational wave approximation (RWA). 
\textbf{b,} Same measurements with the pre-distorted pulse.}
\refstepcounter{EDfig}\label{si_fig9}
\end{figure*}

\begin{figure*}
\centering
\includegraphics[width=2.0\columnwidth]{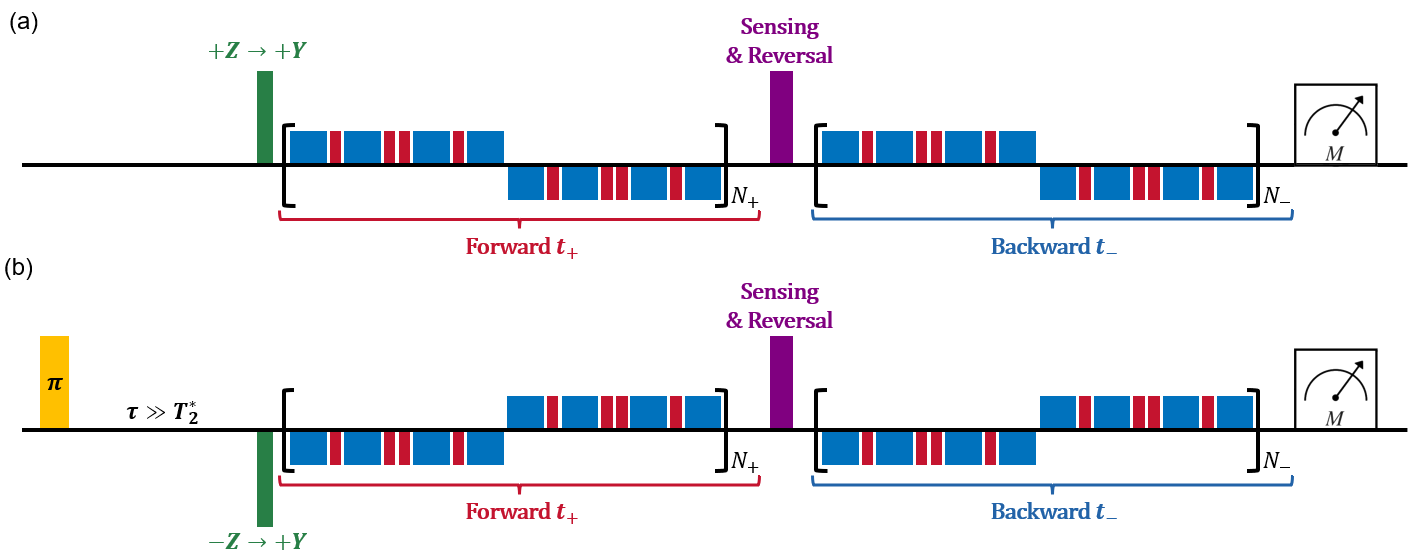}
\caption{\textbf{Robust distance measurements.}
The distances reported in Fig.~\ref{fig3},\ref{fig4} are averaged between the two experiments shown in panel (a) and (b), in order to improve the robustness against systematic errors coming from the (tiny) overlap between the sensing pulse and surrounding Floquet pulses. This strategy works by negating such pulse overlaps, while keeping the engineered Hamiltonian the same. The pulse sequence shown here is slightly different from Fig.~\ref{fig4}(a), as the sensing pulse and the subsequent $\frac{\pi}{2}$ reversal pulse in Fig.~\ref{fig4}(a) are combined into a single pulse (with the same unitary) for simplicity of implementation.
}
\refstepcounter{EDfig}\label{si_fig10}
\end{figure*}

\begin{figure*}
\centering
\includegraphics[width=0.65\textwidth]{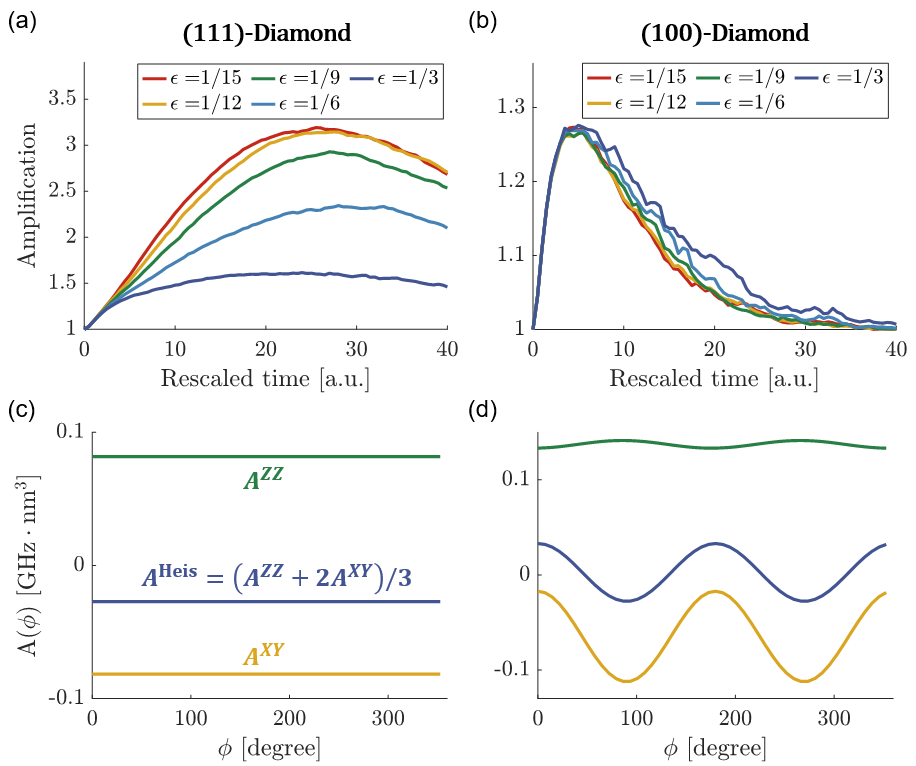}
\caption{\textbf{Possibilities for XY ferromagnetic ordering enhancement.} \textbf{a,} The simulated OAT amplification for Hamiltonians with different distances to the Heisenberg point, in the case of a (111)-oriented diamond. The parameter $\epsilon$ determines the (hypothetical) fraction of time spent in the X, Y, and Z frames, assuming that Floquet sequences are used to engineer the Hamiltonian from its native form. Specifically, the time spent in the X, Y, and Z frames are $\left(\frac{1}{3}-\epsilon,~\frac{1}{3}-\epsilon,~\frac{1}{3}+2\epsilon\right)$, and the horizontal axis is rescaled by $\epsilon$ for the ease of comparison. The amplification is predicted to improve as one engineer the Hamiltonian towards the Heisenberg point, as expected from the XY ferromagnetic ordering mechanism\cite{Block2023}. The simulation is done for a positionally-disordered 2D ensemble without considering experimental imperfections. The sensing angle simulated is $\delta\theta=5^\circ$. \textbf{b,} The same simulation, but for (100)-orientated diamond with the engineered quantization axis $\eta_{\mathrm{eff.}}$ (i.e. the configuration in this work). No improvement is observed as the Hamiltonian is engineered towards the Heisenberg point. \textbf{c,} Dipolar interaction for a (111)-oriented diamond, similar to main text Fig.~\ref{fig1}(c, d). When the Hamiltonian is engineered towards the Heisenberg point from the easy-plane side, the negative $A^{\mathrm{Heis}}$ leads to a tendency for XY ferromagnetic ordering. \textbf{d,} Same as (c), but for (100)-orientated diamond with the engineered quantization axis $\eta_{\mathrm{eff.}}$. The green and yellow curves are the same as main text Fig.~\ref{fig1}(d), repeated here for convenience. The Heisenberg term (blue) shows a mixed sign that leads to magnetic frustration\cite{Vannimenus1977}, which explains the absence of the ordering enhancement in (b).
}
\refstepcounter{EDfig}\label{si_fig11}

\end{figure*}

\clearpage
\newpage

\onecolumngrid

\section*{Supplement Information}
\subsection{Asymmetric echo due to time-reversed mirror symmetry}
In this section, we prove the symmetry of amplification around $t_+ = t_-/2$ under the time-reversed mirror symmetry mentioned in the main text, and further discuss its generality.

According to main text Fig.~\ref{fig4}(e), the time-reversed mirror symmetry flips the sign of time, exchanges the sensing ($\hat{S}$) and measurement ($\hat{M}$) operators, and preserves the initial states ($\ket{\pm\mathrm{Y}}$). This leads to:
\begin{align}
  \chi_{MS}\left(t_+, t_-\right) 
 & \equiv -i\bra{\pm\mathrm{Y}} \lbrack \hat{M}(t_+-t_-), \hat{S}(t_+) \rbrack \ket{\pm\mathrm{Y}}
 \nonumber\\  & = -i\left\{\bra{\pm\mathrm{Y}} \lbrack \hat{S}(t_- - t_+), \hat{M}(-t_+) \rbrack \ket{\pm\mathrm{Y}}\right\}^\star\nonumber\\
    & = -i\bra{\pm\mathrm{Y}} \lbrack \hat{M}(-t_+), \hat{S}(t_- - t_+) \rbrack \ket{\pm\mathrm{Y}}\nonumber\\
  &\equiv\chi_{MS}\left(t_- - t_+, t_-\right),
\label{eq:mirror_symmetry_proof}
\end{align}
where the first line is the Kubo formula\cite{Kubo1957}, the second line comes from the application of the time-reversed mirror symmetry, and the complex conjugate in the second line is a consequence of the anti-unitary\cite{peskin2018introduction} nature of the time-reversal operator. Due to the real value of of the susceptibility $\chi_{MS}$, such complex conjugate is equivalent to a minus sign, which is absorbed in the flipping of the commutator ordering in the third line. Such symmetry guarantees that the amplification takes an extreme value at $t_+ = t_-/2$, and we further emphasize its generality by showing the following observations:
\begin{itemize}
    \item The condition that the time-reversed mirror symmetry exchanges the sensing and measurement operators is not necessary, as long as the sensing and measurement axes are optimized (i.e. chosen to maximize $\chi_{MS}$).
    \item Such time-reversed mirror symmetry applies to a vast majority of Hamiltonians used for entanglement-enhanced sensing.
    \item Although the symmetry does not guarantees a maximum value of amplification at $t_+ = t_-/2$ (as there are concrete examples of local minimum), a maximum is expected in most cases following a heuristic argument.
\end{itemize}

To show that the exchanging of sensing and measurement operators is not necessary, we trace back to the derivation of Eq.~(\ref{eq:mirror_symmetry_proof}). In the case that the mirror reflection ($\mathcal{R}$) does not exchange the sensing and measurement operators, Eq.~(\ref{eq:mirror_symmetry_proof}) generalize to:
\begin{equation}
    \chi_{MS}\left(t_+, t_-\right) = \chi_{S_\mathcal{R} M_\mathcal{R}}\left(t_- - t_+, t_-\right),
\label{eq:generalized_symmetry_SI}
\end{equation}
where $\hat{S}_\mathcal{R}$ and $\hat{M}_\mathcal{R}$ are the mirror reflected sensing and measurement operators. Equation~(\ref{eq:generalized_symmetry_SI}) shows that for any sensing and measurement axes at the forward evolution time $t_+$, there exist a choice of sensing and measurement axes at $t_- - t_+$ that gives the same amplification, and vice versa. Therefore, the amplification under optimized sensing and measurement axes has to obtain equal values:
\begin{equation}
    \chi_{\mathrm{Optimal}}\left(t_+, t_-\right) = \chi_{\mathrm{Optimal}}\left(t_- - t_+, t_-\right),
\label{eq:optimized_symmetry_SI}
\end{equation}
as long as the time-reversed mirror symmetry preserves the initial state. Such symmetry for the OAT Hamiltonian, TAT Hamiltonian, LMG Hamiltonian\cite{LiZeyang2023} ($\hat{H}=\left(\hat{S}^z\right)^2-\abs{S}\hat{S}^y$), and a cubic XYZ Hamiltonian\cite{Zhang2024} ($\hat{H}=\hat{S}^x\hat{S}^y\hat{S}^z + \hat{S}^z\hat{S}^y\hat{S}^x$) are illustrated in Extended Data Fig.~\ref{si_fig2}(a).

To gain further intuition, we simulate the amplification dynamics under these Hamiltonians, for spin ensembles with perfect all-to-all coupling, and dipolar (i.e. $1/r^3$) coupling on a 2D lattice and positionally-disordered 2D plane. As we show in Extended Data Fig.~\ref{si_fig2}(b-d), the predicted symmetry around $t_+ = t_-/2$ is observed in all cases. Remarkably, although not guaranteed by such symmetry, a maximum amplification is observed at $t_+ = t_-/2$ for all cases except the only counter example of OAT in a positionally-disorder ensemble (which can be understood from the splitting of the dimer dephasing-free conditions discussed in the next section and in Extended Data Fig.~\ref{si_fig4}). To understand the generic preference of local maximum instead of local minimum, we give a heuristic argument as follows.

First, we notice that within the gaussian state regime, the amplification is determined solely by $t_-$, as the sensing states with opposite sensing angles $\pm\delta\theta$ are not separated yet during $t_+$. In the gaussian state regime, the amplification continuously improves as $t_-$ increases, until eventually the states go into the highly over-twisted regime (or even thermalized regime in the case of finite range coupled Hamiltonian) where the amplification is degraded. As shown in Extended Data Fig.~\ref{si_fig2}(e-f), asymmetric echo with $t_+ = t_-/2$ allows doubling the amplifying time $t_-$ without encountering such degradation, and is therefore generically expected to give a better amplification. As a warning to readers, we emphasize that this is only a heuristic argument instead of a rigorous statement, as amplification is not necessarily degraded in the highly over-twisted regime (see our complementary work\cite{AmpTheory}, where we show that systems exhibiting collective OAT dynamics continuous to amplify in the highly over-twisted regime). The exact conditions under which a local maximum is expected is an interesting question for future studies.\\

\subsection{Spin dimers: microscopic mechanism for asymmetric echo}
The time-reversed mirror symmetry guarantees an extreme value of amplification at $t_+ = t_-/2$, and a maximum value is expected in most cases. However, to understand exactly why we see a sharp maximum for the TAT Hamiltonian and a minimum for the OAT Hamiltonian in the positionally-disordered case (Extended Data Fig.~\ref{si_fig2}(d)), we need to analyze a microscopic model based on spin dimers that naturally occur in such systems (Extended Data Fig.~\ref{si_fig3}(c)). These observations can be explained by the ``operator phase anti-matching" mechanism discussed in our complementary work\cite{AmpTheory}, which manifests itself as the dephasing and rephasing of the spin dimers. For completeness, the mechanism is explained in the context of dimer dominated systems in detail as below.

As introduced in the main text, we analyze the signal amplification using the formalism of dynamical susceptibility
\begin{align}
    \chi_{MS}(t_+,t_-) &=  -i\bra{\pm\mathrm{Y}} \lbrack \hat{M}(t_+-t_-), \hat{S}(t_+) \rbrack \ket{\pm\mathrm{Y}}
\label{eq:Kubo_formula_SI}
\end{align}
involving the initial state polarized along $\pm\mathrm{Y}$ and the Heisenberg picture evolution of the sensing and measurement operators  $\hat{S}(t),\hat{M}(t)$ (Extended Data Fig.~\ref{si_fig3}(a)). This susceptibility formalism naturally encodes two competing effects. First, as expected from the classical TAT flow, the diagonal components $\chi_{XX}$ and $\chi_{ZZ}$ should obtain non-zero values (Extended Data Fig.~\ref{si_fig3}(b), orange), which contribute to amplification along the $Z+X$ direction. On the other hand, the ubiquitous quantum fluctuations cause additional decay of the Bloch vector beyond classical dynamics\cite{Davis2023,Martin2023}. This is encoded in the decrease of the off-diagonal components $\chi_{XZ}$ and $\chi_{ZX}$ (Extended Data Fig.~\ref{si_fig3}(b), blue) and leads to deamplification.

Motivated by the sharp features in the numerically simulated dynamics (Extended Data Fig.~\ref{si_fig2}(d)) that suggests a diverging energy scale, we consider a toy model involving strongly-coupled pairs of spins - which we call ``spin dimers". For an isolated dimer coupled by a generic XYZ Hamiltonian
\begin{equation}
    \hat{H}_{\mathrm{XYZ}} = J_{\mathrm{D}}\left(g_x \hat{\sigma}^x\hat{\sigma}^x + g_y \hat{\sigma}^y\hat{\sigma}^y + g_z \hat{\sigma}^z\hat{\sigma}^z\right),
\end{equation}
the eigenstates are the Bell states that simultaneously diagonalize $\hat{\sigma}^x\hat{\sigma}^x$, $\hat{\sigma}^y\hat{\sigma}^y$, and $\hat{\sigma}^z\hat{\sigma}^z$:
\begin{align}
    \ket{\psi_-}&\equiv\left(\ket{01}-\ket{10}\right)/\sqrt2, ~~~~~\epsilon_1=J_{\mathrm{D}}\left(-g_x-g_y-g_z\right)\nonumber\\
    \ket{\psi_+}&\equiv\left(\ket{01}+\ket{10}\right)/\sqrt2, ~~~~~\epsilon_2=J_{\mathrm{D}}\left(+g_x+g_y-g_z\right)\nonumber\\
    \ket{\phi_+}&\equiv\left(\ket{00}+\ket{11}\right)/\sqrt2, ~~~~~\epsilon_3=J_{\mathrm{D}}\left(+g_x-g_y+g_z\right)\nonumber\\
    \ket{\phi_-}&\equiv\left(\ket{00}-\ket{11}\right)/\sqrt2, ~~~~~\epsilon_4=J_{\mathrm{D}}\left(-g_x+g_y+g_z\right),
\end{align}
with corresponding eigenenergies $\epsilon_{1,2,3,4}$. Due to the exchange symmetry of the pair coming from global operations and global readout, the entire dynamics is constraint in the symmetric subspace $\{\ket{\psi_+}, \ket{\phi_+}, \ket{\phi_-}\}$, where the total spin operators ($\hat{X}\equiv\hat{\sigma}_1^x+\hat{\sigma}_2^x$, etc) can be expressed as:
\begin{align}
    \hat{X} &= 2\ket{\psi_+}\bra{\phi_+} + \mathrm{h.c.}\nonumber\\
    \hat{Z} &= 2\ket{\phi_+}\bra{\phi_-} + \mathrm{h.c.}
\end{align}
and the initial state as:
\begin{equation}
    \ket{\pm\mathrm{Y}}=\frac{1}{2}\left(\ket{0} \pm i\ket{1}\right)^{\otimes 2} = \frac{\ket{\phi_-}\pm i\ket{\psi_+}}{\sqrt{2}},
\label{eq:initial_state}
\end{equation}
as illustrated in Extended Data Fig.~\ref{si_fig3}(d). In the Heisenberg picture, the evolution of these operators can be described as simple accumulation of phase:
\begin{align}
    \hat{X}\left(t\right) &= 2\mathrm{e}^{i\omega_X t}\ket{\psi_+}\bra{\phi_+} + \mathrm{h.c.}\nonumber\\
    \hat{Z}\left(t\right) &= 2\mathrm{e}^{i\omega_Z t}\ket{\phi_+}\bra{\phi_-} + \mathrm{h.c.}
\label{eq:operator_evolutions}
\end{align}
where $\omega_X = \epsilon_2 - \epsilon_3 = 2J_{\mathrm{D}}\left(g_y - g_z\right)$ and $\omega_Z = \epsilon_3 - \epsilon_4= 2J_{\mathrm{D}}\left(g_x - g_y\right)$ are the dimer spectrum spacing. Intuitively, this allows us to evaluate the diagonal and off-diagonal components of the susceptibility matrix by counting the phase accumulated on the corresponding ``$\Lambda$" and ``ladder" type processes (Fig.~\ref{si_fig3}(d)) that encode Eq.~(\ref{eq:Kubo_formula_SI}). As an example, in the ``$\Lambda$" type process that corresponds to $\chi_{ZZ}$, the sensing operator $\hat{Z}\left(t_+\right)$ acquires a phase $\phi_S = \omega_Z t_+$; and the measurement operator, which happens along the opposite transition, acquires a negative phase $-\phi_M = -\omega_Z \left(t_+ - t_-\right)$. Therefore, the susceptibility is $\chi_{ZZ} \propto \sin{\left(\phi_S - \phi_M\right)} = \sin{\left(\omega_Z t_-\right)}$, contributing to amplification as long as $t_-\neq 0$. Similarly, in the ``ladder" type process that correspond to $\chi_{XZ}$, the sensing and measurement operators acquire phases $\phi_S = \omega_Z t_+$ and $\phi_M = \omega_X \left(t_+ - t_-\right)$, together leading to the susceptibility $\chi_{XZ} \propto \cos{\left(\phi_S + \phi_M\right)}=\cos{\left[\omega_Z t_+ + \omega_X \left(t_+ - t_-\right)\right]}$, corresponding to the expected decay shown in Extended Data Fig.~\ref{si_fig3}(b). Indeed, these intuitions can be confirmed by plugging Eq.~(\ref{eq:initial_state}) and Eq.~(\ref{eq:operator_evolutions}) into Eq.~(\ref{eq:Kubo_formula_SI}), which yields the exact expression of the susceptibility matrix of an isolated dimer:
\begin{align}
    \chi\left(t_+, t_-\right)=\left[
    \begin{array}{cc}
    -\mathrm{sin}\left(\omega_{X}t_-\right) & \mathrm{cos}\left(\omega_{X}t_- - \left(\omega_{X}+\omega_{Z}\right)t_+\right) \\
    -\mathrm{cos}\left(\omega_{Z}t_- - \left(\omega_{X}+\omega_{Z}\right)t_+\right) & \mathrm{sin}\left(\omega_{Z}t_-\right) \\
    \end{array}
    \right].
\label{eq:single_dimer_susceptibility_SI}
\end{align}

In the special case of TAT Hamiltonian studied in this work, the condition $\omega_X = \omega_Z = 2J_{\mathrm{D}}$ leads to further simplification:
\begin{align}
    \chi\left(t_+, t_-\right)=\left[
    \begin{array}{cc}
    -\mathrm{sin}\left(2J_{\mathrm{D}}t_-\right) & \mathrm{cos}\left(2J_{\mathrm{D}}\left(t_- - 2t_+\right)\right) \\
    -\mathrm{cos}\left(2J_{\mathrm{D}}\left(t_- - 2t_+\right)\right) & \mathrm{sin}\left(2J_{\mathrm{D}}t_-\right) \\
    \end{array}
    \right],
    \label{eq:single_dimer_susceptibility_TAT}
\end{align}
where the amplification, which corresponds to sensing along $\frac{Z-X}{\sqrt{2}}$ and measurement along $\frac{Z+X}{\sqrt{2}}$, is
\begin{align}
    \mathrm{Amp} &= \left(\frac{1}{\sqrt{2}},\frac{1}{\sqrt{2}}\right) \chi\left(t_+, t_-\right) \left(-\frac{1}{\sqrt{2}},\frac{1}{\sqrt{2}}\right)^{T}\nonumber\\
    &=\mathrm{sin}\left(2J_{\mathrm{D}}t_-\right) + \mathrm{cos}\left(2J_{\mathrm{D}}\left(t_- - 2t_+\right)\right).
\label{eq:single_dimer_amplification}
\end{align}
As we plot in Extended Data Fig.~\ref{si_fig3}(e), the amplification under the symmetric echo (i.e. $t_+ = t_-$) is $\mathrm{sin}\left(2J_{\mathrm{D}}t_-\right) + \mathrm{cos}\left(2J_{\mathrm{D}}t_-\right)$, whose maximum value is $\sqrt{2}$. In contrast, under the asymmetric echo (i.e. $t_+ = t_-/2$), the amplification is $\mathrm{sin}\left(2J_{\mathrm{D}}t_-\right) + 1$, showing an improved maximum value of $2$. 

In the case of positionally-disorder dimers, the amplification can be calculated by numerically averaging Eq.~(\ref{eq:single_dimer_amplification}) over the distribution of dimer coupling strength $J_{\mathrm{D}}$. Upon such averaging, the sine term contributes to amplification as long as $J_{\mathrm{D}}$ does not average to zero (which is achieved in our 2D sample by engineering the quantization axis); while the cosine term leads to fast dephasing unless $t_- - 2t_+ = 0$. As we plot in Extended Data Fig.~\ref{si_fig3}(f), such dephasing completely washes away any amplification under the symmetric echo; but does not affect the asymmetric echo, which continues to amplify after the disorder averaging. 

Tracing back to the phase accumulation on the ``ladder" type processes, the dephasing free condition $t_- = 2t_+$ fundamentally originates from the anti-matched phase accumulation of the sensing and measurement operators (i.e. $\phi_S + \phi_M = 0$). Although this condition is satisfied simultaneously for $\chi_{XZ}$ and $\chi_{ZX}$ in the case of TAT Hamiltonian, a splitting of the rephasing times is expected for more generic XYZ Hamiltonians. Specifically, according to Eq.~(\ref{eq:single_dimer_susceptibility_SI}), the dephasing free condition for $\chi_{XZ}$ and $\chi_{ZX}$ are $\frac{t_-}{t_+} = 1+\frac{\omega_Z}{\omega_X}$ and $\frac{t_-}{t_+} = 1+\frac{\omega_X}{\omega_Z}$, respectively. To test the understanding based on spin dimers, we check this prediction by comparing the measured peaks of $\Delta X$ and $\Delta Z$ under the TAT Hamiltonian (Extended Data Fig.~\ref{si_fig4}(a)) and another engineered XYZ Hamiltonian (Extended Data Fig.~\ref{si_fig4}(b))
\begin{equation}
    \hat{H}_{\mathrm{XYZ}}=J^{\mathrm{Heis}}\hat{\vec{\sigma}}\cdot\hat{\vec{\sigma}}+\frac{2}{9}J^{\mathrm{Twist}}\left[\left(\hat{\sigma}^z\hat{\sigma}^z-\hat{\sigma}^x\hat{\sigma}^x\right)+\frac{5}{64}\left(\hat{\sigma}^x\hat{\sigma}^x+\hat{\sigma}^z\hat{\sigma}^z-2\hat{\sigma}^y\hat{\sigma}^y\right)\right],
\label{eq:XYZ_Hamiltonian}
\end{equation}
where the dimer spectra is altered. 
We observe a splitting of the $\Delta X$ and $\Delta Z$ peaks under the XYZ Hamiltonian, with locations consistent with the predicted $t_-/t_+$ ratios (1.620 and 2.612) based on the dimer spectra. The local minimum under the OAT Hamiltonian observed in Extended Data Fig.~\ref{si_fig2}(d) can be explained using the same mechanism, but with a more exaggerated peak splitting, as the ``deviation" from the TAT Hamiltonian is even larger.

For interested readers, we provide the Floquet sequence timings for engineering of the XYZ Hamiltonian (Eq.~(\ref{eq:XYZ_Hamiltonian})). Based on similar consideration as in Methods, we chose $t_{\pi,x}=17.67$~ns, $t_{\pi,y}=13.5$~ns, and $\tau=2.75$~ns during the forward evolution $t_+$; and $t_{\pi,x}=14.33$~ns, $t_{\pi,y}=18.5$~ns, and $\tau=5.25$~ns during the backward evolution $t_-$. Here, $t_{\pi,x}$ ($t_{\pi,y}$) is the $\pi$-pulse duration for pulses along the X (Y) axis.\\

\subsection{Contributions to amplification: dimers and regular spins}
Although the simple dimer model captures the locations of optimal amplification, dipolar interactions extend beyond the nearest neighbor coupling in practice. Furthermore, as a consequence of positional disorder in the system, there will naturally be some spins dominated by coupling to their nearest neighbor, and some (more regular) spins coupled more evenly to a few nearby spins. Intuitively, we expect the former to exhibit dynamics dominated by dimer physics, and the latter's dynamics resembles that of a regular lattice. In this section, we try to understand the interplay between dimers and more regular spins, and how they together lead to the predicted amplification of the full dipolar system (Extended Data Fig.~\ref{si_fig5}(e)).

To quantify the difference between dimers and more regular spins, we define the effective coordination number of the $i^{\mathrm{th}}$ spin as:
\begin{equation}
    z_i = \frac{\left(\sum_j \left| J_{ij}\right|\right)^2}{\sum_j J_{ij}^2},
\label{eq:definition_effective_coordination_number}
\end{equation}
where $J_{ij}$ is the coupling between the $i^{\mathrm{th}}$ and $j^{\mathrm{th}}$ spin, and smaller $z_i$ value indicates more dimer-dominated couplings. Using this definition, we calculate the effective coordination number of 200 spins under 1000 randomly sampled positional configurations, and plot its histogram in Extended Data Fig.~\ref{si_fig5}(a). Here, the spins are classified into three groups with low, middle, and high coordination number, each consisting $1/3$ of the total spin number, as indicated by the red, green, and blue shadings on the histogram. We then numerically simulate the dynamics of the 1000 configurations under full dipolar connectivity (and TAT Hamiltonian) without considering experimental imperfections, and plot the average amplification for each group in Extended Data Fig.~\ref{si_fig5}(b-d), where we see:
\begin{itemize}
    \item The low coordination number group shows characteristic dimer peaks along $t_- = 2t_+$, which is not present in the other two groups, as they are not dimer-dominated.
    \item Higher coordination number groups show larger amplification that qualitatively resembles the regular lattice case.
    \item Averaging over the three groups reconstructs the amplification of the full dipolar system (Extended Data Fig.~\ref{si_fig5}(e)). Here, the more regular (i.e. higher coordination number) spins contributed more to the amplification value, but the peak location is still dominated by the dimer physics, due to its sharply peaked features in Extended Data Fig.~\ref{si_fig5}(b).
\end{itemize}

\subsection{Understand the experimental imperfections}
We have seen that without considering experimental imperfections, the pure dimer dynamics leads to an amplification of 19\% (Extended Data Fig.~\ref{si_fig3}(f)), and coupling beyond nearest neighbor improves it to 48\% (Extended Data Fig.~\ref{si_fig5}(e)). However, we only observed an amplification of 6.7\% in the experiment. In this section, we will try to understand the experimental imperfections that leads to this degradation.

As discussed in Methods, relevant experimental imperfections include the imperfect spin polarization (75\%), presence of dynamical on-site disorder ($0.019~\mathrm{MHz}/\sqrt{\mathrm{MHz}}$ around the filter function peak $f_{\mathrm{peak}}=37~\mathrm{MHz}$), and imperfect time-reversal due to the un-reversed Heisenberg term of the Hamiltonian (the static on-site disorder does not have significant effects, as it is decoupled by the Floquet pulse sequence). The effects of these imperfections are studied numerically by turning them off one-by-one in numerical simulations, as shown in Extended Data Fig.~\ref{si_fig6}, where each row (from top to bottom) corresponds to: (1) Experimentally measured data. (2) Our most realistic model including all known imperfections. (3) Turn off the dynamical on-site disorder. (4) Further turn off the imperfect spin polarization, assuming unity polarization. (5) Further turn off the imperfect time-reversal, by reversing the whole Hamiltonian during $t_-$ in the numerical simulation. (6) Same as above, reducing the sensing angle from the experimental value of $15^\circ$ to $5^\circ$ to probe the linear response. (7) Further turn off positional disorder by assuming a perfect 2D lattice geometry, sensing angle is further reduced to $1^\circ$ to guarantee linear response. Here, we have the following main observations:
\begin{itemize}
    \item The most significant limitation is the positional disorder, which reduces the achievable amplification by an order of magnitude. Besides this, the dominant limiting factor is dynamical on-site disorder and imperfect spin polarization, which together qualitatively explain the amount of amplification we observe in the experiment.
    \item In Extended Data Fig.~\ref{si_fig6}(c), we see the predicted symmetry around $t_+ = t_-/2$. Imperfections, including external decoherence (i.e. dynamical on-site disorder), imperfect time-reversal, and dynamics beyond linear response, breaks the exact symmetry.
    \item A large ($>$20~dB) amplification is predicted for the dipolar lattice, suggesting potential application of asymmetric echo in platforms that exhibit such geometry.
\end{itemize}

\subsection{Calculation of the effective Hamiltonian}
To calculate the effective Hamiltonian (Fig.~\ref{fig1}(c,d)) when the quantization axis is changed, we start with the full Hamiltonian for two NV centers coupled by dipole-dipole interaction:
\begin{equation}
    \hat{H} = \hat{H}^{\mathrm{NV}}_1 \left(\vec{B}\right) + \hat{H}^{\mathrm{NV}}_2 \left(\vec{B}\right) + \hat{H}^{\mathrm{Int}},
\label{eq:two_NV_centers}
\end{equation}
where $\hat{H}^{\mathrm{NV}}_1 \left(\vec{B}\right)$ and $\hat{H}^{\mathrm{NV}}_2 \left(\vec{B}\right)$ are the single body terms describing the first and second NV center in the external magnetic field $\vec{B}$, and $\hat{H}^{\mathrm{Int}}$ is the dipolar interaction, given by:
\begin{align}
    &\hat{H}^{\mathrm{NV}} \left(\vec{B}\right) =D \left(\hat{S}^z\right)^2 + \gamma^{\mathrm{NV}}\vec{B}\cdot\hat{\vec{S}}\\
    &\hat{H}^{\mathrm{Int}} = -\frac{J_0}{r^3}\left[3\left(\hat{\vec{S}}_1 \cdot \hat{r}\right)\left(\hat{\vec{S}}_2 \cdot \hat{r}\right) - \hat{\vec{S}}_1 \cdot \hat{\vec{S}}_2\right].
\end{align}
Here, $D = 2\pi\times 2.87~$GHz is the NV zero-field splitting, $\gamma^{\mathrm{NV}} = 2\pi\times 2.8~$MHz/G is the NV spin gyromagnetic ratio, $J_0 = 2\pi\times 52~\mathrm{MHz}\cdot\mathrm{nm}^3$ is the dipolar coupling coefficient, and $\vec{r}$ is the relative position between two NV centers.

In Eq.~(\ref{eq:two_NV_centers}), the energy scale of the single body terms is much larger than the interaction. Therefore, the effective Hamiltonian can be calculated by first diagonalizing the single body terms, and then projecting $\hat{H}^{\mathrm{Int}}$ on the dressed eigenstates $\ket{\widetilde{0}}$ and $\ket{\widetilde{-1}}$ that define the qubit. Moreover, as the energy non-conserving transitions are prohibited, we only need to consider the energy conserving transitions, which include four diagonal components and the flip-flop process:
\begin{align}
    d_{\widetilde{0},\widetilde{0}} &= \bra{\widetilde{0},\widetilde{0}}\hat{H}^{\mathrm{Int}}\ket{\widetilde{0},\widetilde{0}}\nonumber\\
    d_{\widetilde{0},\widetilde{-1}} &= \bra{\widetilde{0},\widetilde{-1}}\hat{H}^{\mathrm{Int}}\ket{\widetilde{0},\widetilde{-1}}\nonumber\\
    d_{\widetilde{-1},\widetilde{0}} &= \bra{\widetilde{-1},\widetilde{0}}\hat{H}^{\mathrm{Int}}\ket{\widetilde{-1},\widetilde{0}}\nonumber\\
    d_{\widetilde{-1},\widetilde{-1}} &= \bra{\widetilde{-1},\widetilde{-1}}\hat{H}^{\mathrm{Int}}\ket{\widetilde{-1},\widetilde{-1}}\nonumber\\
    f &= \bra{\widetilde{0},\widetilde{-1}}\hat{H}^{\mathrm{Int}}\ket{\widetilde{-1},\widetilde{0}}.
\end{align}
Introducing the Pauli operators $\hat{\sigma}_i^x$, $\hat{\sigma}_i^y$, $\hat{\sigma}_i^z$ defined on the qubit space spanned by $\ket{\widetilde{0}}$ and $\ket{\widetilde{-1}}$, these processes contribute to the following effective Hamiltonian:
\begin{align}
    \hat{H}_{\mathrm{eff.}} &= E_0 + J^Z_1 \hat{\sigma}_1^z + J^Z_2 \hat{\sigma}_2^z + J^{ZZ}\hat{\sigma}_1^z \hat{\sigma}_2^z\nonumber\\ &+ \frac{\mathrm{Re}\left[f\right]}{2}\left(\hat{\sigma}_1^x \hat{\sigma}_2^x +\hat{\sigma}_1^y \hat{\sigma}_2^y\right) + \frac{\mathrm{Im}\left[f\right]}{2}\left(\hat{\sigma}_1^x \hat{\sigma}_2^y -\hat{\sigma}_1^y \hat{\sigma}_2^x\right),
\label{eq:initial_effective_Hamiltonian}
\end{align}
with
\begin{align}
    E_0 &\equiv \frac{d_{\widetilde{0},\widetilde{0}} + d_{\widetilde{0},\widetilde{-1}} + d_{\widetilde{-1},\widetilde{0}} + d_{\widetilde{-1},\widetilde{-1}}}{4}\nonumber\\
    J^Z_1 &\equiv \frac{d_{\widetilde{0},\widetilde{0}} + d_{\widetilde{0},\widetilde{-1}} - d_{\widetilde{-1},\widetilde{0}} - d_{\widetilde{-1},\widetilde{-1}}}{4}\nonumber\\
    J^Z_2 &\equiv \frac{d_{\widetilde{0},\widetilde{0}} - d_{\widetilde{0},\widetilde{-1}} + d_{\widetilde{-1},\widetilde{0}} - d_{\widetilde{-1},\widetilde{-1}}}{4}\nonumber\\
    J^{ZZ} &\equiv \frac{d_{\widetilde{0},\widetilde{0}} - d_{\widetilde{0},\widetilde{-1}} - d_{\widetilde{-1},\widetilde{0}} + d_{\widetilde{-1},\widetilde{-1}}}{4}.
\label{eq:J_ZZ_microscopic_expression}
\end{align}
In Eq.~(\ref{eq:initial_effective_Hamiltonian}), the constant energy term $E_0$ has no effect; $J^Z_1 \hat{\sigma}_1^z$ and $J^Z_2 \hat{\sigma}_2^z$ are single body terms that contribute to on-site disorder, which is eventually cancelled by decoupling sequences; and $\mathrm{Im}\left[f\right]$ is guaranteed to be zero due to the exchange symmetry of interaction between two NV centers within the same crystallography orientation group. Therefore, the effective Hamiltonian is guaranteed to simplify to an XXZ form:
\begin{equation}
    \hat{H}_{\mathrm{eff.}} = J^{ZZ}\hat{\sigma}_1^z \hat{\sigma}_2^z + J^{XY}\left(\hat{\sigma}_1^x \hat{\sigma}_2^x + \hat{\sigma}_1^y \hat{\sigma}_2^y\right),
\end{equation}
with $J^{ZZ}$ given in Eq.~(\ref{eq:J_ZZ_microscopic_expression}) and $J^{XY}\equiv\frac{\mathrm{Re}\left[f\right]}{2}$.

To generate Fig.~\ref{fig1}(c,d) in the main text, the above calculations are done numerically.\\

\subsection{Improvements from the pulsed field}
To characterize the improvement of spin polarization with the pulsed field, we compared the Rabi oscillation contrast when the pulsed field is turned on during both initialization and readout, versus it being turned on only during readout (Extended Data Fig.~\ref{si_fig1}(d)). A factor of 4 improvement in the contrast is observed, indicating a factor of 4 improvement in the initial spin polarization.

The effects of the pulsed field on the required averaging time of the experiment is following: First, the initial spin polarization is improved by a factor of 4, leading to an improvement of the signal-to-noise ratio (SNR) by a factor of 4. Second, since the mean field responsible for the twisting dynamics is proportional to the spin polarization, the twisting speed also gets 4 time faster, leading to another factor of 4 improvement in the SNR. Finally, as the initialization is improved by a factor of 4, the readout is also improved by a similar factor because they share the same mechanisms\cite{Doherty2013}. Therefore, in total the SNR of the twisting signal is improved by a factor of $4^3$, which corresponds to a $4^6=4096$ times faster data averaging speed for detection of the twisting dynamics.\\

\subsection{Nuclear spin decoupling}
In this section, we discuss the additional complications coming from the nuclear spin dynamics when an off-axis magnetic field is applied, and the strategy to avoid such complications.

The full Hamiltonian for the two spin system consists of one NV electronic spin and its host $^{15}$N nuclear spin is:
\begin{equation}
    \hat{H} = \hat{H}^{\mathrm{NV}}\left(\vec{B}\right)+\gamma_n \vec{B}\cdot\hat{\vec{I}} + A_{\perp}\left(\hat{J}^x \hat{I}^x+\hat{J}^y \hat{I}^y\right) + A_{\parallel}\hat{J}^z \hat{I}^z,
\label{eq:Full_electron_nuclear_spin_Hamiltonian}
\end{equation}
where $\hat{J}^{x,y,z}$ are NV electronic spin-1 operators, and $\hat{I}^{x,y,z}$ are $^{15}$N nuclear spin-$\frac{1}{2}$ operators, with $z$ axis defined as the native NV axis, and $x,y$ axis perpendicular to $z$. The other parameters includes $\gamma_n = 2\pi\times 0.4316~\mathrm{kHz/G}$ being the gyromagnetic ratio of $^{15}$N, $A_\perp = 2\pi\times 3.65~\mathrm{MHz}$, and $A_\parallel = 2\pi\times 3.03~\mathrm{MHz}$\cite{Felton2009} being the perpendicular and parallel component of the hyperfine tensor.

Since the NV electronic spin energy scale is much larger than the nuclear spin and hyperfine coupling, the hyperfine coupling would never flip the electronic spin. Therefore, we can project the Hamiltonian onto the two electronic eigenstates ($\ket{\widetilde{0}}$ and $\ket{\widetilde{-1}}$, dressed by the $B$ field) that form the NV qubit:
\begin{align}
    \hat{H} = -\frac{E_{\mathrm{NV}}}{2}\hat{\sigma}^z &+ \gamma_n \vec{B}\cdot\hat{\vec{I}} + \ket{\widetilde{0}}\bra{\widetilde{0}}\left[A_\perp {\langle\hat{J}^x\rangle}_{\ket{\widetilde{0}}}\hat{I}^x + A_\perp {\langle\hat{J}^y\rangle}_{\ket{\widetilde{0}}}\hat{I}^y + A_\parallel {\langle\hat{J}^z\rangle}_{\ket{\widetilde{0}}}\hat{I}^z\right]\nonumber\\
    &+\ket{\widetilde{-1}}\bra{\widetilde{-1}}\left[A_\perp {\langle\hat{J}^x\rangle}_{\ket{\widetilde{-1}}}\hat{I}^x + A_\perp {\langle\hat{J}^y\rangle}_{\ket{\widetilde{-1}}}\hat{I}^y + A_\parallel {\langle\hat{J}^z\rangle}_{\ket{\widetilde{-1}}}\hat{I}^z\right],
\end{align}
where $\hat{\sigma}^z$ is the spin-$\frac{1}{2}$ Pauli operator defined on this qubit, and ${\langle\hat{J}^{x,y,z}\rangle}_{\ket{\widetilde{0}},\ket{\widetilde{-1}}}$ are the expectation values of spin operators on these two states. Going into the rotating frame with respect to the NV frequency $E_{\mathrm{NV}}$, and pick the definition of $x$ and $y$ such that the external $B$ field points in the XZ plane, the Hamiltonian can be further simplifies as:
\begin{equation}
    \hat{H} = \frac{\hat{\sigma}^z}{2}\left(a\hat{I}^z + b\hat{I}^x\right) + c\hat{I}^z + d\hat{I}^x,
\label{eq:important_nuclear_coupling}
\end{equation}
where the parameters $a,b,c,d$ are given by:
\begin{align}
    a &=A_\parallel\left[{\langle\hat{J}^z\rangle}_{\ket{\widetilde{0}}} - {\langle\hat{J}^z\rangle}_{\ket{\widetilde{-1}}}\right]\nonumber\\
    b &=A_\perp\left[{\langle\hat{J}^x\rangle}_{\ket{\widetilde{0}}} - {\langle\hat{J}^x\rangle}_{\ket{\widetilde{-1}}}\right]\nonumber\\
    c &=\gamma_n B_z + A_\parallel\frac{{\langle\hat{J}^z\rangle}_{\ket{\widetilde{0}}} + {\langle\hat{J}^z\rangle}_{\ket{\widetilde{-1}}}}{2}\nonumber\\
    d &=\gamma_n B_x + A_\perp\frac{{\langle\hat{J}^x\rangle}_{\ket{\widetilde{0}}} + {\langle\hat{J}^x\rangle}_{\ket{\widetilde{-1}}}}{2}.
\end{align}

In the case of the external $B$ field parallel to the native NV axis, we have ${\langle\hat{\vec{J}}\rangle}_{\ket{\widetilde{0}}}=\vec{0}$, and ${\langle\hat{\vec{J}}\rangle}_{\ket{\widetilde{-1}}}$ pointing along the NV axis. Therefore, we have $b=d=0$, and the Hamiltonian simplifies to $\hat{H} = \frac{a}{2}\hat{\sigma}^z\hat{I}^z + c\hat{I}^z$. This Hamiltonian commutes with $\hat{I}^z$, so that the nuclear spin polarization is static and therefore produces a predominantly static field to the NV center. Such a static field is then cancelled by decoupling sequences, in exactly the same way as on-site disorder gets cancelled.

However, when the external $B$ field has a non-zero off-axis component, things become more complicated. Here, the term $a\hat{I}^z+b\hat{I}^x$ in Eq.~(\ref{eq:important_nuclear_coupling}) generates the on-site field, but the field may not be static because $a\hat{I}^z+b\hat{I}^x$ does not commute with the Hamiltonian. In general, this leads to complicated dynamics of the NV electronic spin such as electron-spin-echo-envelope-modulation (ESEEM) effect\cite{Rowan1965,Ohno2012}; but further simplification is possible under the assumption that a fast decoupling sequence is applied on the NV electronic spins, as in this work.

Under the assumption that a fast (i.e. much faster than the nuclear spin dynamics) decoupling sequence is applied on the NV electronic spin, the coupling between the electronic spin and the nuclear spin get significantly suppressed, as the electronic spin is nearly decoupled from the slowly varying field generated by the nuclear spin, due to the presence of the decoupling sequence. As a consequence, the nuclear spin dynamics is dominated by the term $c\hat{I}^z + d\hat{I}^x$ in Eq.~(\ref{eq:important_nuclear_coupling}). Therefore, it is convenient to change into the eigenbasis of this term by defining:
\begin{align}
    \hat{\tilde{I}}^z &\equiv \frac{c\hat{I}^z + d\hat{I}^x}{\sqrt{c^2 + d^2}}\nonumber\\
    \hat{\tilde{I}}^x &\equiv \frac{c\hat{I}^x - d\hat{I}^z}{\sqrt{c^2 + d^2}}.
\end{align}
The Hamiltonian is rewritten in this basis as:
\begin{equation}
    \hat{H} = \frac{ac+bd}{2\sqrt{c^2+d^2}}\hat{\sigma}^z\hat{\tilde{I}}^z + \frac{bc-ad}{2\sqrt{c^2+d^2}}\hat{\sigma}^z\hat{\tilde{I}}^x + \sqrt{c^2+d^2}\hat{\tilde{I}}^z.
\end{equation}
Going into the interaction picture (i.e. rotation frame) with respective to the term $\sqrt{c^2+d^2}\hat{\tilde{I}}^z$, the Hamiltonian is transformed into:
\begin{equation}
    \hat{H} = \frac{ac+bd}{2\sqrt{c^2+d^2}}\hat{\sigma}^z\hat{\tilde{I}}^z + \frac{bc-ad}{2\sqrt{c^2+d^2}}\hat{\sigma}^z\left(\hat{\tilde{I}}^x \cos{\Omega_N t} - \hat{\tilde{I}}^y \sin{\Omega_N t}\right),
\label{eq:resulting_NV_nuclear_coupling}
\end{equation}
where $\Omega_N\equiv\sqrt{c^2+d^2}$ is the nuclear spin precession frequency. Here, the first term has no time dependence and therefore gets cancelled by the decoupling sequences in the same way as on-site disorder; but the second term has an explicit time dependence, which could get in synchronize with the filter function of the decoupling sequence and lead to periodic entangling and disentangling dynamics between the NV electronic spin and the nuclear spin.

To study the conditions for such synchronization, it is worth noticing that the base frequency of the filter function is always $1/T_{\mathrm{Floquet}}$, where $T_{\mathrm{Floquet}}$ is the Floquet period (i.e. the duration of the repeated pulse sequence). Therefore, synchronization generically happens when the Floquet period matches the nuclear spin precession period, which should be avoided when choosing the timings of pulse sequences. Furthermore, to achieve the best nuclear spin decoupling, it is more desirable to set $T_{\mathrm{Floquet}}=\frac{k}{m} T_{\mathrm{Nuc}}$, where $T_{\mathrm{Nuc}}$ is the nuclear spin precession period, and $m\nmid k$ are small integers, as the coupling to the nuclear spin can be cancelled completely within time $kT_{\mathrm{Nuc}}$.

In this work, the B field is set to $B_x = 143~$G and $B_z = 877~$G, which leads to the parameters $a = 2\pi\times 1.837~\mathrm{MHz}$, $b = 2\pi\times \left(-4.258\right)~\mathrm{MHz}$, $c = 2\pi\times \left(-1.132\right)~\mathrm{MHz}$, and $d = 2\pi\times \left(-0.076\right)~\mathrm{MHz}$. The corresponding nuclear spin precession period is therefore $T_{\mathrm{Nuc}}=\frac{2\pi}{\sqrt{c^2+d^2}}=881~$ns, and the TAT engineering sequence is (approximatly) synchronized to it when the $\pi$-pulse duration is set to $t_\pi = 24~$ns (Extended Data Fig.~\ref{si_fig7}), where the Floquet period is $T_{\mathrm{Floquet}}=864~$ns. The Floquet periods in TAT and time-reversed TAT measurements (Fig.~\ref{fig3},\ref{fig4}) are both $T_{\mathrm{Floquet}}=432~$ns, roughly satisfying $T_{\mathrm{Floquet}}=\frac{1}{2} T_{\mathrm{Nuc}}$. The Floquet period in the peak splitting measurement (Extended Data Fig.~\ref{si_fig4}(b)) is $T_{\mathrm{Floquet}}=576~$ns, roughly satisfying $T_{\mathrm{Floquet}}=\frac{2}{3} T_{\mathrm{Nuc}}$.\\

\subsection{Possibilities for XY ferromagnetic ordering enhancement}
It is predicted that if an XXZ Hamiltonian is engineered towards the Heisenberg point from the easy-plane side, one can improve the squeezing/amplification relying on the XY ferromagnetic ordering mechanism\cite{Block2023}. We numerically analyzed this possibility (Extended Data Fig.~\ref{si_fig11}) by simulating the amplification under dipolar OAT dynamics (i.e. Eq.~(\ref{eq:Heisenberg_Ising_decomposition}) in main text) for Hamiltonians with different distances to the Heisenberg point, for both (111)-oriented diamonds and (100)-oriented diamonds with the engineered quantization axis $\eta_{\mathrm{eff.}}$ (i.e. the configuration in this work). Although the quantization axes are normal to the 2D plane in both cases, we emphasize that they are still different configurations. An intuitive understanding is that because NV centers are spin-1 particles with a finite zero field splitting, the picture in Fig.~\ref{fig1}(b) of main text only approximately applies to the states $\ket{\widetilde{0}}$ and $\ket{\widetilde{-1}}$, but not to the third state $\ket{\widetilde{+1}}$. The existence of the third state then complicates the calculation, and gives an effective Hamiltonian in the $\{\ket{\widetilde{0}}, \ket{\widetilde{-1}}\}$ subspace that is different from that of a (111)-oriented diamond. In these simulations, we assume the engineering of Hamiltonian by Floquet sequences, with the fraction of time spent in X, Y, and Z frames being $\left(\frac{1}{3}-\epsilon,~\frac{1}{3}-\epsilon,~\frac{1}{3}+2\epsilon\right)$, where the parameter $\epsilon$ indicates the distance to the Heisenberg point. The simulation is done without considering experimental imperfections, except the positional disorder and finite 2D layer thickness.

Based on the results shown in Extended Data Fig.~\ref{si_fig11}, we see that the amplification could be improved by engineering the Hamiltonian towards the Heisenberg point for (111)-oriented diamonds, but not for (100)-oriented diamonds with the engineered quantization axis. To explain why this strategy fails to work for (100)-oriented diamonds, we compare the dipolar coupling for the (111) and (100) cases in Extended Data Fig.~\ref{si_fig11}(c, d). In contrast to the (111) case where the negative Heisenberg term (blue trace) leads to a tendency for XY ferromagnetic ordering (when the Hamiltonian is engineered towards the Heisenberg point from the easy-plane side), the Heisenberg term in the (100) case exhibits a mixture of ferromagnetic coupling and anti-ferromagnetic coupling, and therefore has no tendency for ferromagnetic ordering. Therefore, improvement of the amplification by the ordering mechanism requires a (111)-oriented diamond, and we expect that advances in the growth of such diamonds\cite{Hughes2024} to lead to significantly larger signal amplification.

\end{document}